\documentclass[onefignum,onetabnum]{siamart171218}

\usepackage{lipsum}
\usepackage{amsfonts}
\usepackage{graphicx}
\usepackage{subcaption}
\usepackage{tikz}
\usepackage{quantikz}
\usepackage{epstopdf}
\usepackage{algorithmic}
\usepackage{tabularx}
\ifpdf
  \DeclareGraphicsExtensions{.eps,.pdf,.png,.jpg}
\else
  \DeclareGraphicsExtensions{.eps}
\fi

\newsiamremark{remark}{Remark}
\newsiamremark{hypothesis}{Hypothesis}
\crefname{hypothesis}{Hypothesis}{Hypotheses}
\newsiamthm{claim}{Claim}

\headers{Solving nonlinear PDEs via a hybrid Newton Method using QLSS}{M. Mandelt Buxadé, S. Langer, and P. Bekemeyer}

\title{Solving nonlinear Partial Differential Equations via a hybrid Newton Method using Quantum Linear System Solver\thanks{Submitted to the editors 19.12.2025.
\funding{This project was made possible by the DLR Quantum Computing Initiative and the Federal Ministry for Economic Affairs and Climate Action;
\url{qci.dlr.de/projects/toquaflics}}}}

\author{Maximilian Mandelt Buxadé\thanks{Institute for Aerodynamics and Flow Technology, German Aerospace Center (DLR), Lilienthalplatz 7, 38108
Braunschweig, Germany 
  (\email{maximilian.mandeltbuxade@dlr.de}, \email{stefan.langer@dlr.de}, \email{philipp.bekemeyer@dlr.de}).}
\and Stefan Langer\footnotemark[2]
\and Philipp Bekemeyer\footnotemark[2]}

\usepackage{amsopn}

\makeatletter
\newcommand*{\addFileDependency}[1]{
  \typeout{(#1)}
  \@addtofilelist{#1}
  \IfFileExists{#1}{}{\typeout{No file #1.}}
}
\makeatother

\ifpdf
\hypersetup{
  pdftitle={Solving nonlinear Partial Differential Equations via a hybrid Newton Method using Quantum Linear System Solver},
  pdfauthor={M. Mandelt Buxadé, S. Langer and P. Bekemeyer}
}
\fi

\begin{document}

\maketitle

\begin{abstract}
  To approximate solutions of complex nonlinear partial differential equations remains a computational challenge, especially for sets of equations relevant in industry, such as Euler or Navier-Stokes equations. 
  Even the most sophisticated computational fluid dynamic algorithms coupled with powerful supercomputers can not find approximate solutions for several design challenges in both adequate time and scale-resolving accuracy. One difficulty arises from solving high dimensional, strongly nonlinear partial differential equations, such as the Navier-Stokes equations, which capture the underlying physics. For nearly all classical algorithms, methods closely related to Newton's method are used to approximate a solution to the problem.
  Approximately solving the large-scale linear systems of equations occurring in this iterative scheme is generally a main contributor to the total computational complexity.
  In this paper a new quantum linear system solver supporting Newton's classical method to solve nonlinear partial differential equations is introduced.
  We present a new variant of the HHL algorithm, requiring less apriori information regarding the eigenvalues of the corresponding matrix. We apply this quantum linear system solver in a hybrid quantum-classical fashion to solve nonlinear partial differential equations. Moreover, a resource estimation for advanced use-cases of practical relevance is provided. 
  Our results demonstrate how quantum computation may improve existing classical methodologies for solving nonlinear partial differential equations. This approach provides another promising application of quantum computers and presents a possible way forward for handling nonlinearities on inherently linear quantum systems. 
\end{abstract}



\section{Introduction}
In research, as well as in industry solving boundary-value problems (BVP) using certain partial differential equations~(PDEs) is an integral part in various areas. Often, the PDEs to be solved are nonlinear and it is open if the BVP has a known unique solution. Therefore, typically numerical schemes are required.
Computational Fluid Dynamics~(CFD) is an important, established tool for such simulations targeting flows \cite{AndersonCFD.1995}. For high Reynolds number flows typically the Reynolds averaged Navier-Stokes~(RANS) equations are solved in combination with a turbulence model to keep computational cost at bay \cite{Slotnick.2014}. However, many years of experience in applying these methods have shown that the accuracy of predictions decreases significantly, for example towards the borders of an aircraft's flight envelope, and that scale-resolving methods should be used instead~\cite{Slotnick.2014}.
Unfortunately, even the most sophisticated algorithms cannot, in general, compute such a scale-resolving approximation of a solution for an advanced use-case, for example a complete aircraft, within an computational complexity acceptable for even the most powerful supercomputers. Moreover, to further spread the adoption of CFD for improving engineering designs, for example reducing the drag and hence operating cost of an aircraft, not just one, but a significant number of such scale-resolving calculations is necessary~\cite{Slotnick.2014}.\\
Most of the computational time comes from approximately solving the large-scale, non-linear systems of the equations obtained after discretization. Typically, approaches based on implicit Runge-Kutta-methods which can be interpreted as a regularized Newton method are used to approximate a solution~\cite{La:1401,La:18}. Hence, the computational complexity of solving the large-scale linear systems of equations, occurring in this iterative scheme, is a main contributor to the total complexity~\cite{inexactNewtonMethod}. Due to the scaling with the problem size, which is linear in the best case, simply increasing computational resources is not a sustainable way forward and improved methods must be developed.\\
In 2009 the first algorithm was proposed, which can solve linear systems using a quantum computer with an exponential advantage over previous methods \cite{Harrow.2009}. To obtain exponential advantage when compared with classical methods various conditions need to be satisfied \cite{Aaronson.2015, Harrow.2009}. The so called Harrow-Hassidim-Lloyd~(HHL) algorithm was further improved (e.g. \cite{Cao.2012, Childs.2017HHLImproved}) and applied to various use-cases, see for example \cite{Clader.2013HHLusecase, Jing.2024}. To choose certain parameters of the algorithm some prior knowledge of the eigenvalues of the problem matrix is demanded. To overcome this, Cao et al.~\cite{Cao.2012} proposed a variant, but it introduces a high computational overhead or inaccuracies for certain eigenvalues \cite{arcsinCost.Haner.31.05.2018}. In the context of inverted black box state preparation, a proposal of new techniques, useful for writing inverted amplitudes of bit-encoded quantum states, was made \cite{Wang.2022}. While having the same functionality as the changes proposed by Cao et al.~\cite{Cao.2012} they do not have the same drawbacks. The circuit proposed uses multiplication operators \cite{parent2017improved} and a routine to compare the size of two values saved in quantum registers \cite{Cuccaro.2004}. This form of black box state preparation based on comparisons was first introduced in \cite{Sanders.2019}.
 The combination of Newton's method or a variation of it with a quantum linear system solver~(QLSS), such as the HHL algorithm, was studied for machine learning \cite{Rebentrost.2019, Wossnig.19.10.2017}, electrical engineering \cite{ElKhatib.2024} and even general sparse systems of nonlinear equations \cite{Xue.17.09.2021}. Using HHL in hybrid iterative schemes has been proposed in the field of CFD for the lid driven cavity test case \cite{Lapworth.2022} and for linear problems \cite{Bharadwaj.2023}.\\
The hybrid methodology presents a different approach of handling nonlinearities compared to global linearization methods, which use the quantum system for the full algorithm. Most known are the Carleman \cite{carleman1932application} or the Koopman spectral linearization \cite{Koopman1931, KoopmanNeumann1932}. The Carleman linearization has been applied to sparse matrices (see \cite{Krovi.2023} for example) and problems with a limited dissipation-nonlinearity ratio \cite{LiuCarlemanDissipativeNonlinear.2021}. It was also compared to some less known linear embeddings \cite{Engel.2021Carlemann}. Additionally, The Koopman operator has been applied to variety of dynamical systems \cite{GiannakisKoopmannDynamicSystems.2022, Joseph.2020KoopmanDynamicSystems}. Both methods require the implementation of a matrix, which scales suboptimal with the problem size \cite{Shi_KoopmanVSCarleman.29.10.2023}. Additionally, the application of the so called Schrödingerization \cite{Jin.16.08.2023MaxwellSchrodingerisation, JinSchrodingerization2023} was demonstrated on PDEs related to CFD \cite{Hu.15.03.2024, Meng.2023, Sato.2024}. These global linearizations are challenged by highly nonlinear functions, as faced in CFD areas.\\
In this paper we present a new variant of the HHL algorithm, applying ideas from black box state preparation. This variant has reduced assumptions about the eigenvalues of the matrix to be inverted. To accelerate the procedure of solving nonlinear PDEs, we propose a way a QLSS can be used in Newton's method to approximately solve the inner linear systems. Both the QLSS and the hybrid method are applied in a simulated environment and as a classical algorithm, allowing to study its characteristics for various accuracies and equations. Different aspects of using a quantum algorithm within the classical method are discussed with respect to potential speed up, accuracy and scaling. The obtained result shows a potential use of quantum computers in the area of CFD, even when strong nonlinearities are present. The proposed methodology can be used in a wider context, to solve BVP in a variety of areas.\\
While hybrid Newton methods have been proposed before, it was done so in other areas of application. The previous proposals also focus on the original HHL method and mostly restrict themselves to theoretical studies. Here a unique numerical study applying a variant of the HHL is provided. The ability of the hybrid method to solve nonlinear PDEs is demonstrated, underlining the advantage of this method in the area of CFD compared to other approaches. \\
 The outline of the paper is as follows. In Section \ref{methodology} the methodology will be explained. Details of Newton's method are shown in Section \ref{sec:Meth_Newton} which is then followed by a description of the HHL and its variant in \ref{sec:QLSS}. Specific details regarding implementation and a classical version of the QLSS are given in Section \ref{sec:implementation} and Section \ref{sec:ModelQLSS} respectively. Afterwards, in Section \ref{simulations}, simulations are shown for a variety of both linear and nonlinear problems. We discuss the obtained results and additionally put them in context with practical considerations in Section \ref{discussion}. Additionally, estimates for industry relevant use cases are presented. An outlook for future work and summarizing the findings presented here is given in Section \ref{conclusion}.

\section{Methodology}
\label{methodology}

\begin{figure}
    \centering
    \includegraphics[width=0.8\linewidth]{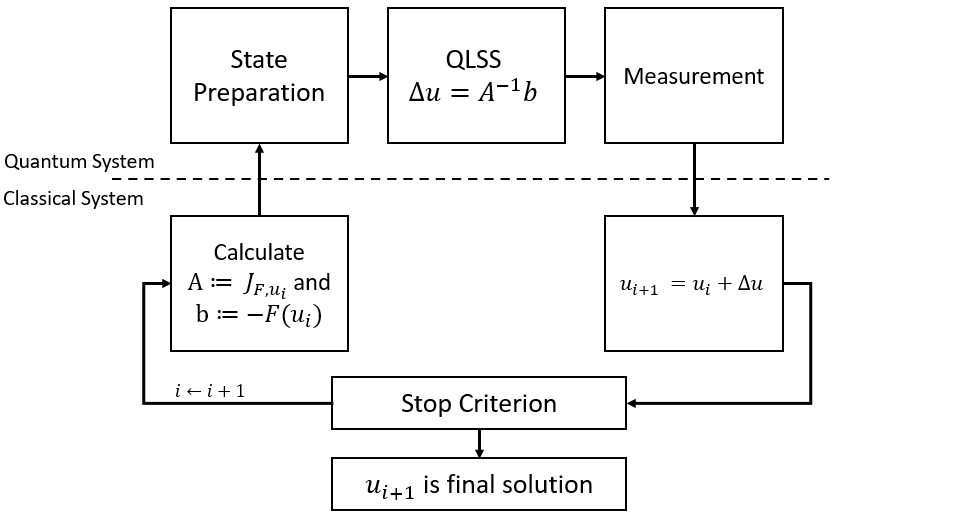}
    \caption{The general workflow of a quantum Newton method. Here, $J_{F, u_i}$ denotes the Jacobian for the function $F$ of the equation $F(u) = 0$. The quantum system provides the iterative step $\Delta u$.}
    \label{fig:generalIdea}
\end{figure}
The method pursued herein relies on combining Newton's method with a linear system solver realized on a quantum computer. This general idea was presented in various contexts~\cite{ElKhatib.2024, Rebentrost.2019, Wossnig.19.10.2017, Xue.17.09.2021}. A blueprint of the workflow is given in Figure \ref{fig:generalIdea}.
In the following subsections both parts of the proposed method are discussed. Section \ref{sec:Meth_Newton} introduces the  classical and widely known Newton's method. Followed by an explanation of the quantum computing part of the hybrid method, which solves the linear system in Section \ref{sec:QLSS}. The novel QLSS is one of the main differences to previous quantum Newton methods (QNM). In Section \ref{sec:implementation} we elaborate on the specific implementation of the QLSS used here and provide details on the simulating process. Finally, in Section \ref{sec:ModelQLSS} a classical version of the algorithm is presented, which will greatly benefit the numerical study.
\subsection{Newton's Method}
\label{sec:Meth_Newton}
Newton's method is a widely known mathematical method to solve nonlinear equations. A in-depth description of it and some variants are described by \cite{La:18}. Herein we restrict ourselves to the use of the most simple form of the algorithm, as a proof of concept, but for more industry-related use-cases different variations play a key-role \cite{La:18, Trottenberg.2007Multigrid}.\\
Consider the problems
\begin{equation}
    F(x) = y,
\end{equation}
where $F: X\rightarrow Y$ is a nonlinear function, $x\in X$ the unknown variable to be determined and $y\in Y$. It is generally assumed that $X$ and $Y$ are Hilbert or Banach spaces. The iterative scheme is described by the equation $x_{i+1} = x_i - F'(x_i)^{-1} F(x_i)$.\\
In the context of $X=\mathbb{R}^d$, $d=2$ or $3$ we often write $u\in X$ instead of $x$ and the operator $F'$ is equivalent to applying the Jacobian matrix $J_F$. So the iteration steps can be written as
\begin{align}
    \Delta u &= J_F^{-1}(u_i)(-F(u_i)+y),\label{NewtonLinSys}\\
    u_{i+1} &= u_i + \Delta u,
\end{align}\\
where often $y=0$. Here the task at hand becomes solving the linear system in Equation \ref{NewtonLinSys}. A well-studied method, often used in practice, is the so called Gauß-Seidel method, see \cite{Trottenberg.2007Multigrid} for details.
In application the iterative Newton scheme is stopped according to some criterion, which either limits the number of iterations or stops once $\Delta u$ is smaller than some threshold.

\subsection{Quantum Linear System Solver}
\label{sec:QLSS}
It is important to note that a QLSS does not solve the classical linear system, but a related problem. Here we follow the work by Morrell et al. \cite{Morrell.2023StepbyStepHHL}. Let $A$ be a hermitian matrix of size $N\times N$ with only positive eigenvalues $\lambda_i$, i.e. $\lambda_i > 0$ for all $i=1,...,N$. We may assume without loss of generality that $A$ is hermitian, because for non-hermitian matrices we exchange $A$ for 
$\begin{pmatrix}
    0&  A\\
    A^\dagger&  0
\end{pmatrix}$
and adapt $b$ and $x$ accordingly, see the work of Harrow et al. \cite{Harrow.2009} for details. A quantum algorithm is called a QLSS, if for any prepared state 
\begin{equation}
    |b\rangle:= \sum_{i=0}^{N} \beta_i |i\rangle = \sum_{i=0}^{N} b_i |u_i\rangle
\end{equation}
the method calculates the state
\begin{equation}
\label{SolutionState}
    |x\rangle := \sum_{i=0}^{N} \frac{b_i}{\lambda_i}|u_i\rangle,
\end{equation}
where $|u_i\rangle$ denote the eigenvectors of their respective eigenvalue $\lambda_i$. This is related to the classical problem of $Ax = b$, but contains normalization.\\
\subsubsection{HHL Algorithm}
The first method, solving such a quantum linear system and preparing the state $|x\rangle$, was proposed in 2009 by Harrow et al. \cite{Harrow.2009} and is often called HHL algorithm. This algorithm is explained in great detail by Morrell et al. \cite{Morrell.2023StepbyStepHHL} and will only be summarized here. 
The algorithm can be separated into four major parts and acts on three quantum registers. The register $B$ contains the state $|b\rangle$. The clock register $C$ and the ancilla register $Anc$ are both initialized in the zero state $|0\rangle$. As mentioned before, we assume that the state $|b\rangle$ is already prepared, so we do not count state preparation as one of the algorithm's steps. Nevertheless, it is an important step of any procedure containing this algorithm.\\
The HHL algorithm starts by applying quantum phase estimation (QPE). This widely used procedure, as mentioned by \cite{Nielsen.2012.QCandQI}, was first presented in \cite{Kitaev.1995QPE}. The goal of QPE here is to approximately write the eigenvalues $\lambda_i$, $i=1,...,N$ via a computational encoding into the clock Register $C$. This is achieved by approximating the phase of the eigenvalues of the unitary operator $e^{iAt}$. Therefore the circuit transforms the initial state, such that
\begin{equation}
    \operatorname{QPE}(|b\rangle_B|0\rangle_C|0\rangle_{Anc}) = \sum_{i=0}^{N}b_i|u_i\rangle_B|\tilde{\lambda}_i\rangle_C|0\rangle_{Anc},
\end{equation}
with $\tilde{\lambda}_i$ being a bit-encoded approximation of the eigenvalue $\lambda_i$. Here, depending on number of qubits $m$ of register $C$ and parameter $k$ the accuracy of the QPE is determined by the approximation 
\begin{equation}
\label{lambdaApprox}
    \lambda_i \approx \tilde{\lambda}_i =2^{m-k}\tilde{\lambda}_{i,m-k}+...+2^{-k}\tilde{\lambda}_{i,-k}\quad,\tilde{\lambda}_{i,j}\in\{0,1\}, j=m-k,...,-k.
\end{equation}
To motivate the next step, recall that the final state the algorithm aims to achieve in register $B$ is given by Equation \ref{SolutionState}. Observe that the eigenvalues are contained inverted in the amplitude of the quantum states. To achieve this, the next step performs a rotation $R_Y$ controlled by the encoded eigenvalues. The rotation itself is given by
\begin{equation}
\label{Ry}
    R_Y(\theta):= \begin{pmatrix}
    \cos{\frac{\theta}{2}}&  -\sin{\frac{\theta}{2}}\\
    \sin{\frac{\theta}{2}}&  \cos{\frac{\theta}{2}}
\end{pmatrix}.
\end{equation}
This rotation aims to transfer the computationally encoded eigenvalues to their inverted self in the amplitude of the ancilla register, which translates to
\begin{equation}
\label{eq:controlledRotation}
    \operatorname{cR_Y}(|\lambda\rangle|0\rangle) := |\lambda\rangle \left(\sqrt{1-\frac{c^2}{\lambda^2}}|0\rangle + \frac{c}{\lambda} |1\rangle\right), 
\end{equation}
with $c$ being a constant freely chosen. One chooses $c$ to maximize the probability of measuring the ancilla register $Anc$ in state $|1\rangle_{Anc}$. One can now observe that the desired result of the inverted eigenvalue $\lambda^{-1}$ is contained in the amplitude, if the ancilla is in state $|1\rangle$. Therefore, the circuit continues by measuring the ancilla register and post-selects only the measurements with the correct outcome. The state of the post-selected outcomes is then given by
\begin{equation}
    |\psi\rangle = \sum_{i=0}^{N}b_i|u_i\rangle_B |\tilde{\lambda}_i\rangle_C \frac{c}{\tilde{\lambda}_i}|1\rangle_{Anc},
\end{equation}
omitting a normalization factor.\\
In a final step the registers $C$ and $B$ are disentangled by uncomputation, i.e. performing inverse QPE. Afterwards, the final state becomes
\begin{equation}
    |x\rangle_B|0\rangle_C|1\rangle_{Anc}
\end{equation}
again omitting normalization factors.\\
To  obtain the full exponential advantage over classical methods, this algorithm poses some conditions that must be fulfilled. These conditions are mentioned in the original paper \cite{Harrow.2009}, but are more clearly stated in the work of Aaronson \cite{Aaronson.2015}, and can be summarized as follows:
\begin{itemize}
    \item One needs to initialize $b$ efficiently,
    \item $e^{iAt}$ as part of the QPE must be efficiently applied by the quantum computer,
    \item $A$ must be well-conditioned,
    \item one can not read out the full state $x$.
\end{itemize}
It is generally considered that if $A$ is sparse, $e^{-iAt}$ can be applied efficiently \cite{Berry_2015}.
\subsubsection{Proposed QLSS}
To understand why a variation of the HHL algorithm is preferable for the proposed QNM, consider the performed controlled rotation. The used rotation gate is given by Equation \ref{Ry} and it acts on a state $|0\rangle$ as
\begin{align}
    R_Y(\theta)(|0\rangle) &= \cos{\frac{\theta}{2}} |0\rangle + \sin{\frac{\theta}{2}} |1\rangle \\
    &=\sqrt{1-\sin^2\left(\frac{\theta}{2}\right)} |0\rangle + \sin{\frac{\theta}{2}} |1\rangle,
\end{align}
where the second part gives a better intuition how one obtains Equation (\ref{eq:controlledRotation}).\\
Now consider the eigenvalue $\tilde{\lambda}$, similar to Equation \ref{lambdaApprox}, but for simplicity assuming only integer encoding, i.e. $k=0$, written in binary form 
\begin{equation}
\tilde{\lambda}= 2^m\tilde{\lambda}_{m} + ...+2^0\tilde{\lambda}_{0}, \quad\tilde{\lambda}_{i}\in\{0,1\}.
\end{equation} 
Let $\theta_i$ be the angle with which the ancilla in the circuit is rotated, if and only if $\tilde{\lambda}_i$ is $1$. Then one needs to choose $\theta_i$ such that
\begin{equation}
    \sum_{i=0}^m \tilde{\lambda}_i R_Y(\theta_i)|0\rangle =K|0\rangle + \frac{c}{\sum_{i=0}^m 2^i\tilde{\lambda}_i}|1\rangle
\end{equation}
 for some $K$. In other words the goal is to approximate $\frac{c}{\tilde{\lambda}}$ using $\sum_{i=0}^m \tilde{\lambda}_i \sin{\frac{\theta_i}{2}}$. Unfortunately, this is not achievable for general $\tilde{\lambda}$.\\
To illustrate this, consider the following example. For the eigenvalue $\tilde{\lambda} = 1 = 2^0\tilde{\lambda}_0$ we aquire 
\begin{equation}
    \tilde{\lambda}_0 \sin{\frac{\theta_0}{2}} = \frac{c}{\tilde{\lambda}_0} \implies \theta_0=2\arcsin{\frac{c}{\tilde{\lambda}_0^2}}=2\arcsin{c}.
\end{equation} 
Similarly, $\tilde{\lambda} = 2 = 2^1 \tilde{\lambda}_1$ results in $\theta_1 =2\arcsin{\frac{c}{2\tilde{\lambda}_1^2}} = 2\arcsin{\frac{c}{2}}$. Considering $\tilde{\lambda} = 3 = 2^0\tilde{\lambda}_0 + 2^1 \tilde{\lambda}_1$, the equation 
\begin{equation}
    \tilde{\lambda}_0\sin{\frac{\theta_0}{2}}+ \tilde{\lambda}_1\sin{\frac{\theta_1}{2}} = \frac{c}{2^0\tilde{\lambda}_0 + 2^1\tilde{\lambda}_1}
\end{equation}
should hold, but inserting the previously obtained angles $\theta_0$ and $\theta_1$ the left hand side becomes \begin{equation}
    \frac{c}{2^0\tilde{\lambda}_0} + \frac{c}{2^1 \tilde{\lambda}_1} = \frac{c}{1} + \frac{c}{2} \neq \frac{c}{3}.
\end{equation}
Hence a general selection of $\theta_i$ is impossible.\\
Therefore, in the literature it is considered beneficial to transform $|\tilde{\lambda}\rangle$ to $|\tilde{\lambda}^{-1}\rangle$, as for example by Cao et al. \cite{Cao.2012}, and then perform the conditional rotation. This does work for small eigenvalues, for which $1/\tilde{\lambda}$ is large, because then $\frac{1}{\tilde{\lambda}}\approx \arcsin{\frac{1}{\tilde{\lambda}}}$. But for this scheme to work for all possible eigenvalues one would need to use quantum arithmetic to compute $|\arcsin\frac{1}{\tilde{\lambda}}\rangle$, which introduces an impractical overhead \cite{arcsinCost.Haner.31.05.2018}. One could also calculate additional information of the eigenvalues to make the choice of the angles possible. Since the hybrid application here would make it necessary to do these calculations in each iteration this would introduce an unwanted overhead too.\\
There is another approach avoiding conditional rotations and approximating $\frac{1}{\tilde{\lambda}}$ in the amplitude directly from the computational encoding of the eigenvalue using comparison of two registers \cite{Wang.2022}. We can use this method, originally proposed for black-box state preparation, to obtain a more practical HHL variant, where less knowledge of the eigenvalues is necessary.\\
The general procedure of the method, as shown for general inverse state preparation in the work of Wang et al. \cite{Wang.2022}, is similar to the original HHL method, starting with a QPE using a register containing vector $b$ and a register initialized as $|0\rangle$. Additionally, as can be observed in Figure \ref{fig:WangCirc}, a $m$-sized register $M_1$, a $2m$-sized register $M_2$ and, apart from implementation based ancilla, one ancilla qubit are needed. Therefore, after QPE we obtain the state
\begin{equation}
    |\psi_1\rangle = \left(\sum_{i=0}^N b_i|u_i\rangle|\tilde{\lambda}_i\rangle\right)|0\rangle_{M_1}|0\rangle_{M_2} |0\rangle.
\end{equation}
Afterwards, the conditional rotation is replaced by a different strategy to encode the inverted eigenvalues into the amplitude. The register $M_1$ is now prepared into a uniform superposition $\frac{1}{\sqrt{2^m}}\sum_{j=0}^{2^m}|j\rangle$, where $j$ is written in binary form using $m$ bits. For this, Hadamard matrices $H$ on each of the respective $m$ qubits must be applied. In other words, the tensor product $H^{\otimes m}$ is used as follows
\begin{equation}
    \left(\sum_{i=0}^N b_i|u_i\rangle|\tilde{\lambda}_i\rangle\right)H^{\otimes m}|0\rangle_{M_1}|0\rangle_{M_2} |0\rangle=\sum_{i=0}^N b_i|u_i\rangle|\tilde{\lambda}_i\rangle\frac{1}{\sqrt{2^m}}\sum_{j=0}^{2^m}|j\rangle_{M_1}|0\rangle_{M_2}|0\rangle.
\end{equation}
This state is then multiplied via arithmetic with the prepared eigenvalues and the result is stored in register $M_2$, so
\begin{equation}
    |\psi_2\rangle=\left(\sum_{i=0}^N b_i|u_i\rangle|\tilde{\lambda}_i\rangle\frac{1}{\sqrt{2^m}}\sum_{j=0}^{2^m}|j\rangle|j\cdot \tilde{\lambda}_i\rangle\right) \otimes|0\rangle.
\end{equation}
The goal now is to use this state to encode the inverted eigenvalues in the amplitude and obtain $\frac{b_i}{\tilde{\lambda}_i}$. For this, introduce the comparison operator, which compares two registers $a$ and $b$, and flips an ancilla qubit to $|1\rangle$, if and only if $a>b$. To understand why the comparison is able to approximate the inverse consider the fact that $x\cdot \frac{1}{x}=1$, so
\begin{equation}
    y < \frac{1}{x} \implies x\cdot y<1 \text{ and } y > \frac{1}{x} \implies x\cdot y>1.
\end{equation}
Now, comparing the states $|j\cdot\tilde{\lambda}\rangle$, $j=0,...,2^m$ with $|2^{m-k}\rangle$ and counting the cases of the ancilla being $|0\rangle$ gives a good approximation for $\frac{1}{\tilde{\lambda}}$.
This is, again as in the original algorithm, followed by the uncomputation and finally the measurement. The post-selection is done on all outcomes where $M_1$, $M_2$ and the ancilla are $|0\rangle$. The success probability is given by
\begin{equation}
    P_{succ} = \left(2^{-\frac{m}{2}}\left|\left|\left(1/\tilde{\lambda}_i\right)_{i=0,...,N}\right|\right|_2\right)^2/n,
\end{equation}
but can be boosted by amplitude amplification \cite{Brassard_2002}. The final result is then identical to the HHL algorithm, ignoring ancilla qubits and renormalization factors. To obtain the full exponential runtime improvement over classical methods the same conditions as for the HHL algorithm must be fulfilled.
\subsubsection{Error Bounds and Asymptotic Scaling}
\label{sec:errorBoundsandAsymScaling}
Now that the variant of the HHL algorithm has been defined, let us determine the obtained error and the expected scaling of this method. Given, the method is based on the work by Wang et al. \cite{Wang.2022} and simply applied to solving linear systems, we refer to that work for a detailed error analysis. Given a binary encoding with $m$ bits (see Equation \ref{lambdaApprox}), we use $\lceil m/2 \rceil$ bits for integer bits and $k=\lfloor m/2 \rfloor$ for fractal bits, rounding in favor of integer bits. This bounds the eigenvalue approximation by $2^{-(\frac{m}{2})}\leq\tilde{\lambda}<2^{\frac{m}{2}}$ and therefore also the inversion of it with the same bounds. The error for both the QPE and the inversion is given by $\varepsilon\sim 2^{-(m/2)}$, as long as $2^{-m/2}\leq \lambda \leq 2^{m/2}$ is satisfied. Note, one could use two different accuracies for the QPE and for the inversion, see \cite{Wang.2022}, but for simplicity here the same accuracy for both is assumed.\\
A detailed resource cost of the algorithm of course depends on the practical implementation and underlying hardware, but some general remarks can be made. The input register size scales with $n=\log_2(N)$, whereas $N\times N=2^n\times2^n$ is the size of the matrix $A$. As stated in \cite{Wang.2022} the additional register sizes scale with $O(\log_2(\varepsilon^{-1}))=O(m)$. The circuit depth of QPE generally scales linearly with $\frac{1}{\varepsilon}$ \cite{Kitaev.1995QPE, Nielsen.2012.QCandQI} and is the main contributor to the total depth, i.e. the scaling of the other algorithmic steps added in the variant has a lower scaling with respect to $\varepsilon$. This results in the same scaling of $\operatorname{poly}(\kappa, n, \frac{1}{\varepsilon})$ as the original HHL algorithm \cite{Harrow.2009}, with $\kappa$ being the condition number of $A$.

\subsection{Implementation}
\label{sec:implementation}
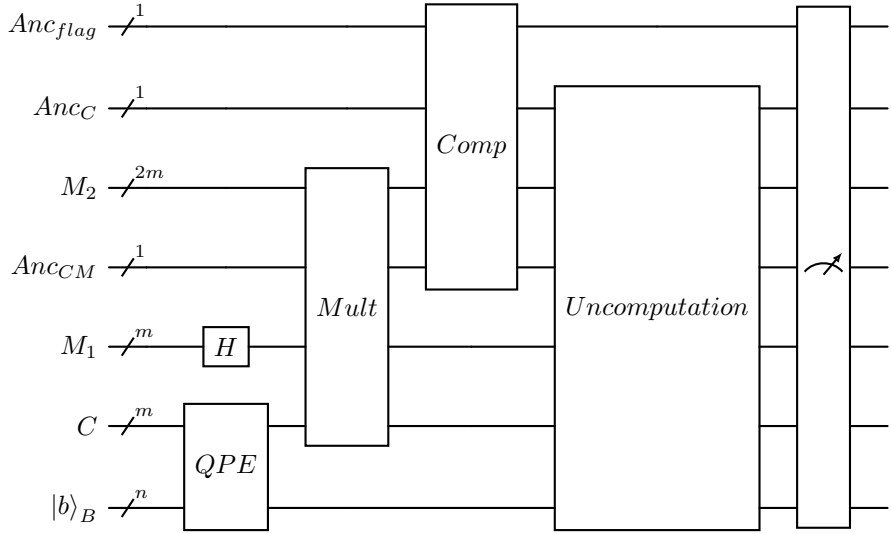
\begin{figure}
    \centering
    \begin{quantikz}[transparent]
    \lstick{$Anc_{flag}$}& \qwbundle{1} &&&\gate[4]{Comp}&& \meter[7]{}& \\
    \lstick{$Anc_C$}& \qwbundle{1}      &&&&\gate[6]{Uncomputation}&&\\
    \lstick{$M_2$}& \qwbundle{2m}       &&\gate[4]{Mult}&&&&\\
    \lstick{$Anc_{CM}$}& \qwbundle{1}   &&&&&&\\
    \lstick{$M_1$}& \qwbundle{m}        &\gate{H}&&&&&\\
    \lstick{$C$}& \qwbundle{m}          &\gate[2]{QPE}&&&&&\\
    \lstick{$\ket{b}_B$} & \qwbundle{n} &&&&&&
    \end{quantikz}
    \caption{Quantum circuit of the implemented QLSS. Here, $Anc_{C,M}$ is the ancilla for the multiplication. This register is also used together with register $Anc_C$ for the comparison. The uncomputation undoes all the operations, except the comparison.}
    \label{fig:WangCirc}
\end{figure}

Performing numerical analysis utilizing the proposed hybrid methods requires the implementation of the proposed QLSS in the form of a gate-based quantum circuit. A high-level depiction of the circuit can be seen in Figure \ref{fig:WangCirc}. In the following section we will provide an overview of the components and their respective implementation.\\
The QPE is a widely studied algorithm for which many variations and improvements for different applications exist, see the work of Ni et al. \cite{Ni.2023LowDepthQPE} for examples. For our studies it suffices to use the general implementation, as can be found in the book of Nielsen and Chuang \cite{Nielsen.2012.QCandQI}. This contains the implementation of controlled Hamiltonian simulations and the quantum Fourier transform.
For the multiplication step of QLSS a circuit implementation is provided by Parent et al. \cite{Parent.12.06.2017}. The result of the multiplication is stored out-of-place, contrary to replacing a register, since the circuit requires to be invertible. Further, the result needs $2m$ qubits to store a multiplication of two $m$ qubit registers to prevent an overflow. The method uses an adder circuit, which is implemented according to Cuccaro et al. \cite{Cuccaro.2004}. The adder scales in depth with $O(\log(\varepsilon^{-1}))$ for given accuracy $\varepsilon$, the multiplication scales with $O(\log^2(\varepsilon^{-1}))$. Therefore the asymptotic depth of the multiplication is negligible compared to the depth of the QPE. Additionally, the comparison is based on the implemented adder as well, see \cite{Cuccaro.2004}. Due to the fact, that the method compares the multiplication result always to $2^m$ the comparison can be significantly simplified, resulting in one instead of $m$ ancilla qubits. The uncomputation before measurement and postselection is straightforward by applying all necessary operations in reverse. Using all this building blocks one obtains the aforementioned schematic circuit as shown in Figure \ref{fig:WangCirc}.\\
Since the matrices in general are not hermitian, the linear system will be modified as mentioned in Section \ref{sec:QLSS}. Note that for a two-dimensional grid of $N\times N$ of a PDE the matrix that needs to be inverted is of size $N^2 \times N^2$. Given that this matrix is not hermitian the system size increases to $2N^2 \times 2N^2$. While classically this would introduce a prohibitive overhead in computation, on the quantum system this leads to just one additional qubit, underlining the scaling powers of quantum computing. This implementation is used to run simulations in the open-source Python framework Qiskit \cite{JavadiAbhari.14.05.2024_Qiskit}.\\
For the current proof of concept and to obtain a first understanding of the capabilities of the presented method, both state preparation and a full read-out are assumed. Additionally, some advancements for various components of this circuit exist, but do not improve the asymptotic scaling. Therefore we restricted ourselves to basic implementations and simply refer to improvements in the following. A more developed multiplication circuit can be found together with the basic implementation \cite{Parent.12.06.2017}. An improvement in the gate count for the adder can be achieved following Gidney \cite{Gidney.2018Adder}. While again the asymptotic scaling is not changed by these methods and the overall cost is dominated by the QPE a implementation on real hardware should try to minimize the gate count as much as possible and make use of any advancements.
\subsection{Model QLSS}
\label{sec:ModelQLSS}
Due to the limited simulation capabilities, to obtain valuable insights into the scaling of the proposed algorithm an additional classical version of the algorithm was implemented. The classical version, which will be denoted model QLSS, does not simulate the full quantum dynamics, but rather focuses on the approximation steps of the algorithm. The model QLSS therefore first approximates the eigenvalues to given $m$-bit accuracy, where half of the available bits are used for the integer encoding and the other half for the fractal encoding. The approximated eigenvalues are then each multiplied with all possible numbers between $0$ and $2^m-1$. The results of these multiplications are then compared to $2^m$. Now, counting the number of comparisons for which $2^m$ is bigger, gives us the approximated inverted eigenvalues in the same fashion as the quantum method. While this does not directly translate to full simulations of a quantum computer, it enables investigations of the noise-free performance of the proposed algorithm for up to hundreds of qubits without the computational overhead of a quantum simulator.\\
Beyond that, even the model QLSS uses considerable amounts of computational resources. One quantum multiplication and comparison classically turn into $2^m$ multiplications followed by $2^m$ comparisons per eigenvalue. This underlines the power of quantum computing and how superpositions and entanglement allow for strong scaling advantages.  

\section{Results}
\label{simulations}
In this section we apply the proposed methodology to a variety of problems. First, the QLSS will be applied to linear problems and compared to the classical method in \ref{sim:LinProb}. Afterwards we employ the full methodology of the QNM to a nonlinear Poisson equation and to the Burgers equation in \ref{sim:QNM}.
\subsection{Linear Problems}
\label{sim:LinProb}
\begin{figure}[t]
    \centering
    \includegraphics[width=0.45\linewidth]{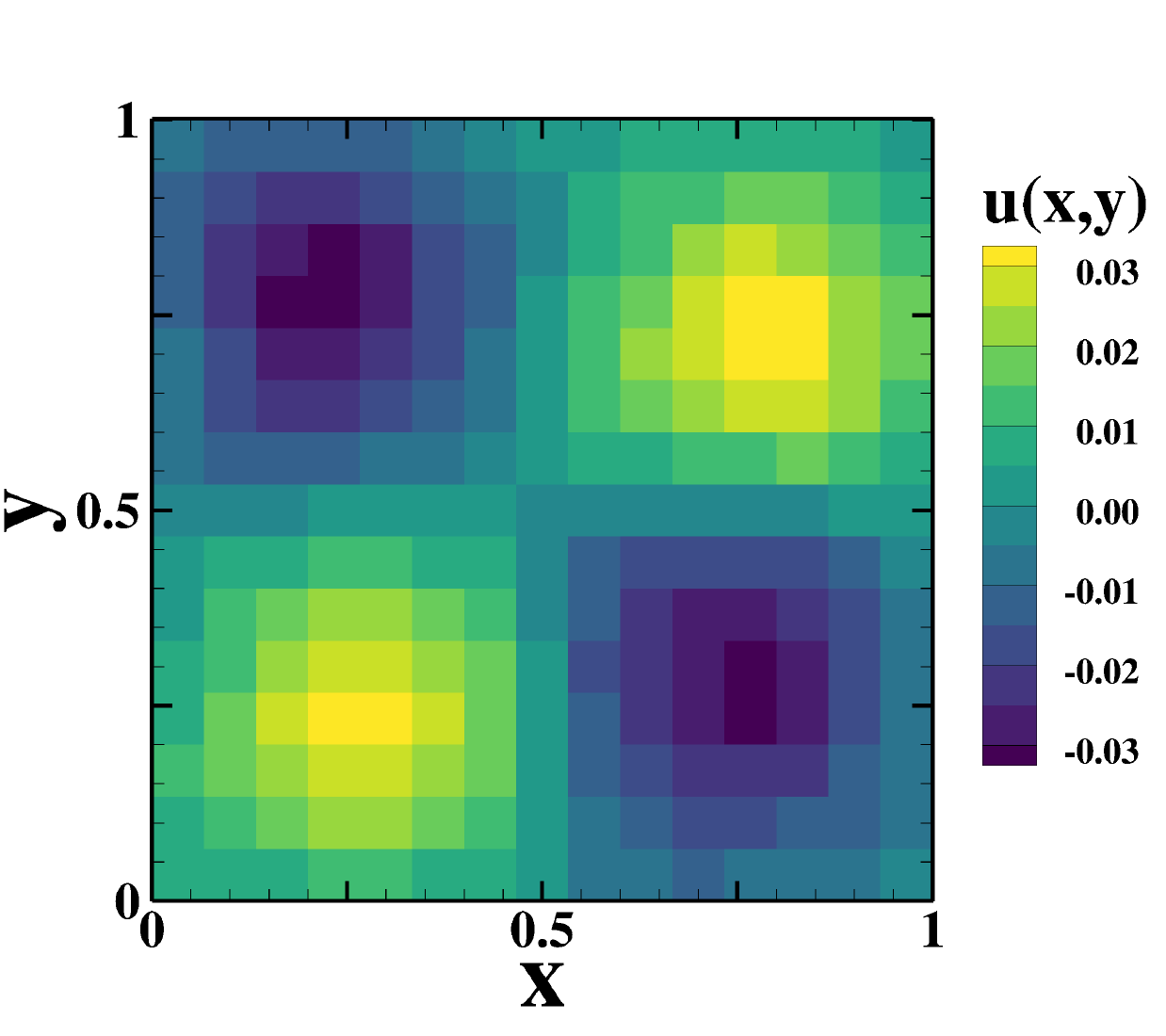}
    \caption{The calculated normalized solution of the stated Advection Diffusion equation for $N=16$. The solution is obtained via Qiskit simulation of the QLSS with $m=2$.}
    \label{fig:vortexWangSol}
\end{figure}

\begin{figure}[htb]
\begin{minipage}[t]{0.45\linewidth}
\includegraphics[width=\linewidth]{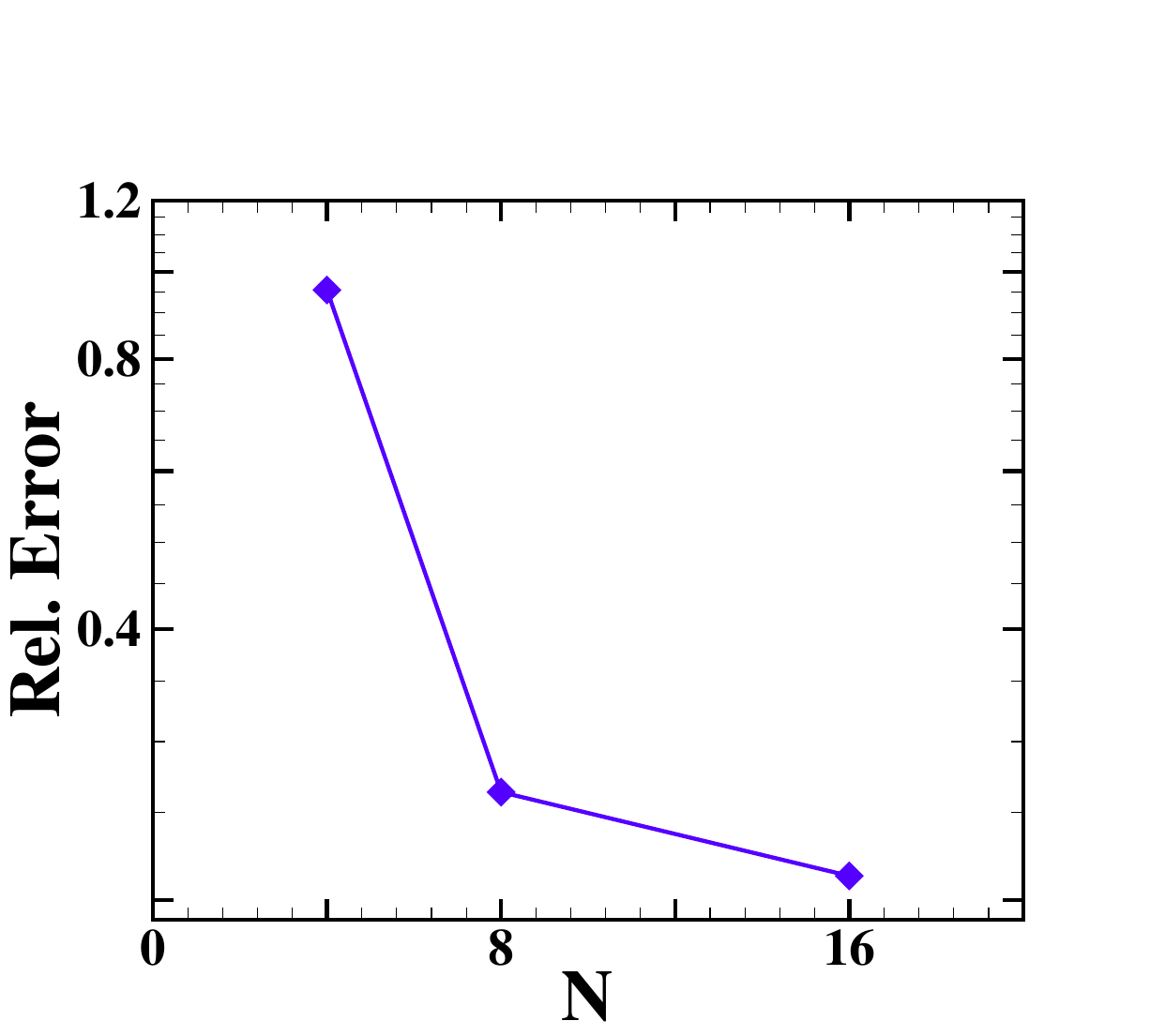}
\caption{Relative error of the calculated Advection Diffusion equation solution using an $m=2$ QLSS with different discretizations resulting in $N\times N$ sized linear systems.}
\label{fig:errorPlotExp1}
\end{minipage}
\hfill
\begin{minipage}[t]{0.45\linewidth}
\includegraphics[width=\linewidth]{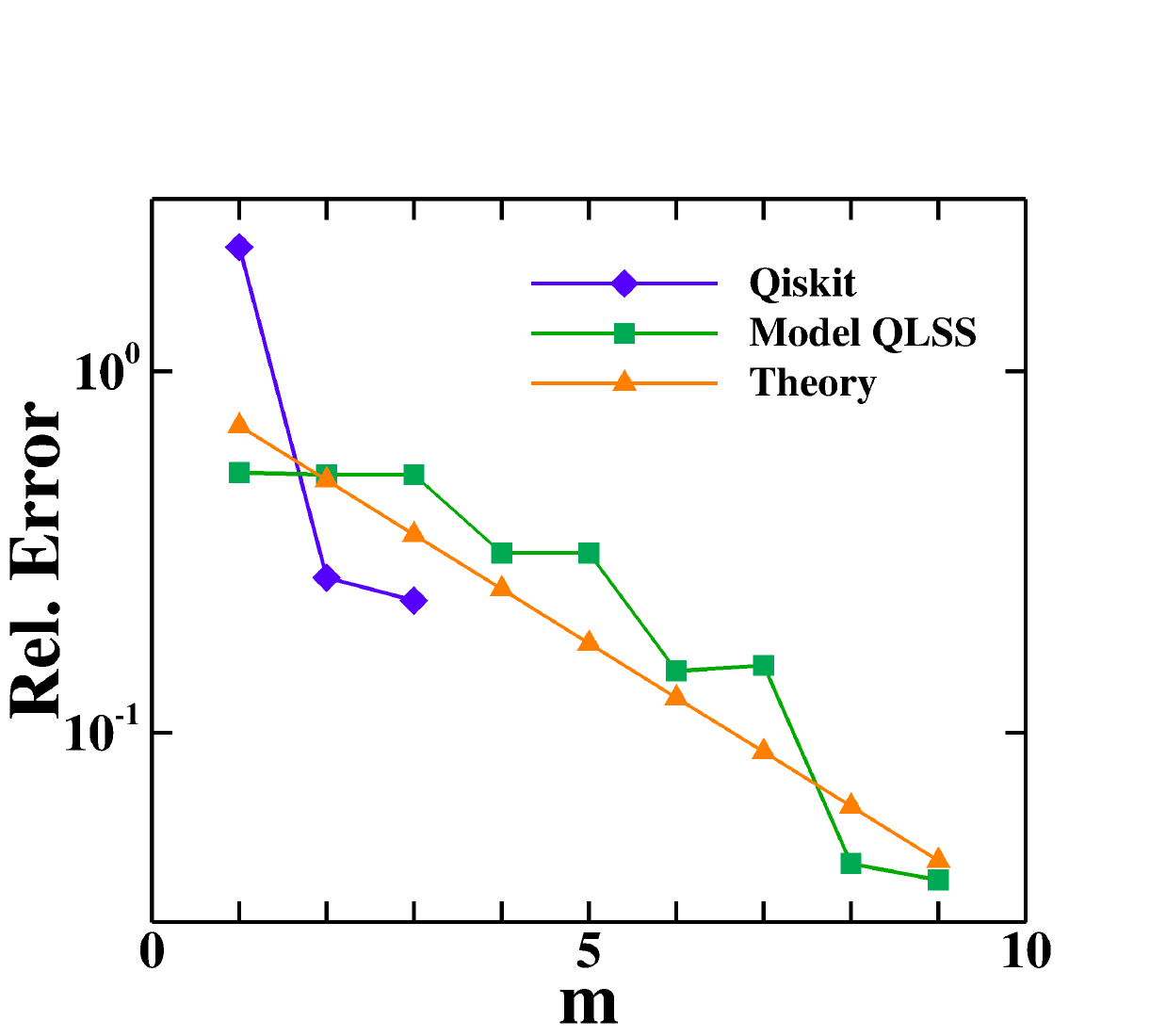}
\caption{Relative error for various choices of $m$, the number of qubits for the QPE for solving the Advection Diffusion equation, $N=8$. Theoretically the error in the approximations scales with $\varepsilon = 2^{-\frac{m}{2}}$ for $m$ qubits.}
\label{fig:exp3}
\end{minipage}%
\end{figure}

\begin{figure}[tb]
\begin{subfigure}[t]{0.45\textwidth}
\centering
\includegraphics[width=\linewidth]{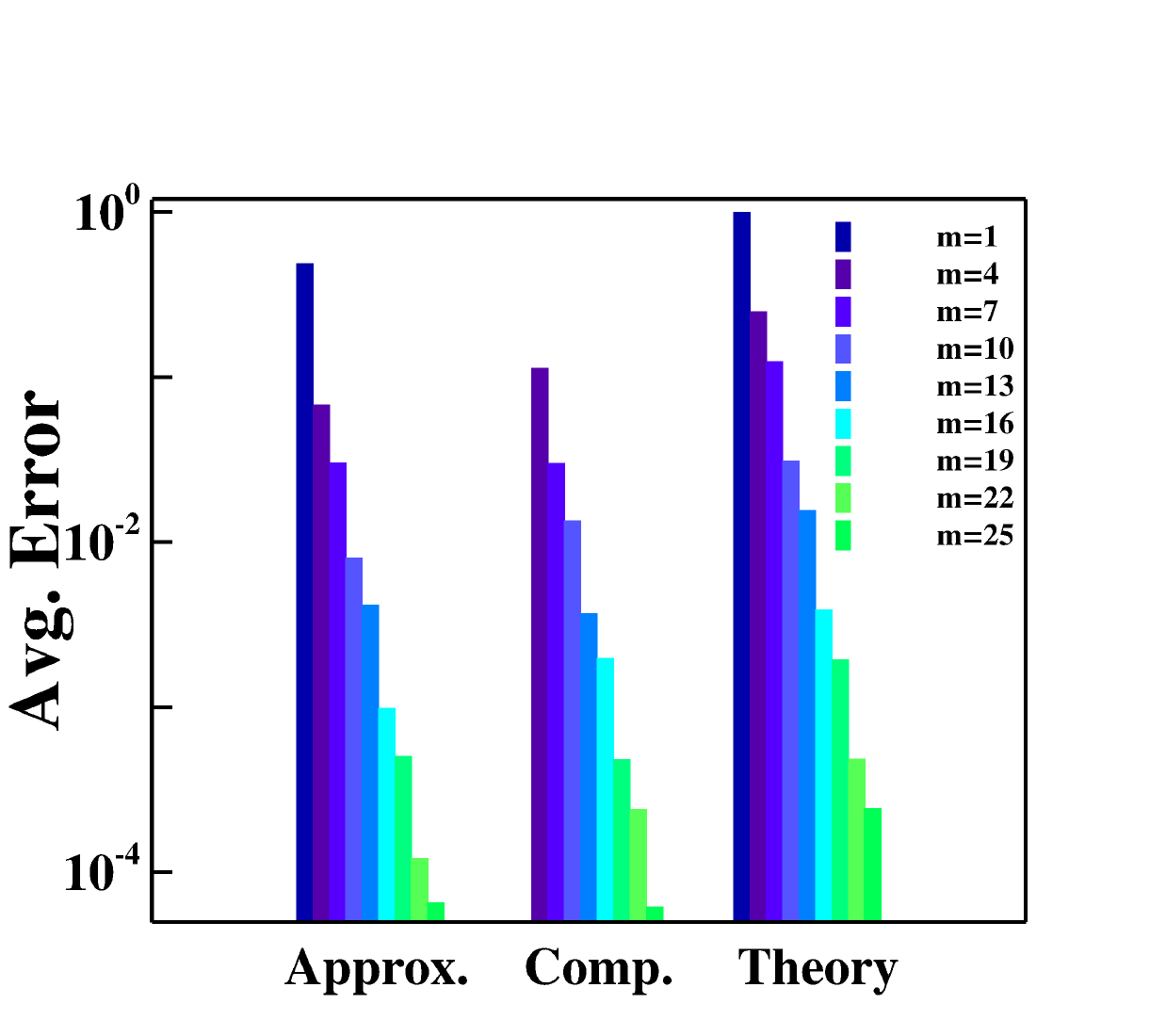}
\caption{Breakdown of the two main steps of the model QLSS. Here 'Approximation' denotes the error of the $m$-bit approximation of the eigenvalues and 'Comparison' denotes the error in obtaining the inverted eigenvalues by the comparison procedure.}
\label{fig:ErrorBySource}
\end{subfigure}%
\hfill
\begin{subfigure}[t]{0.45\textwidth}
\centering
\includegraphics[width=\linewidth]{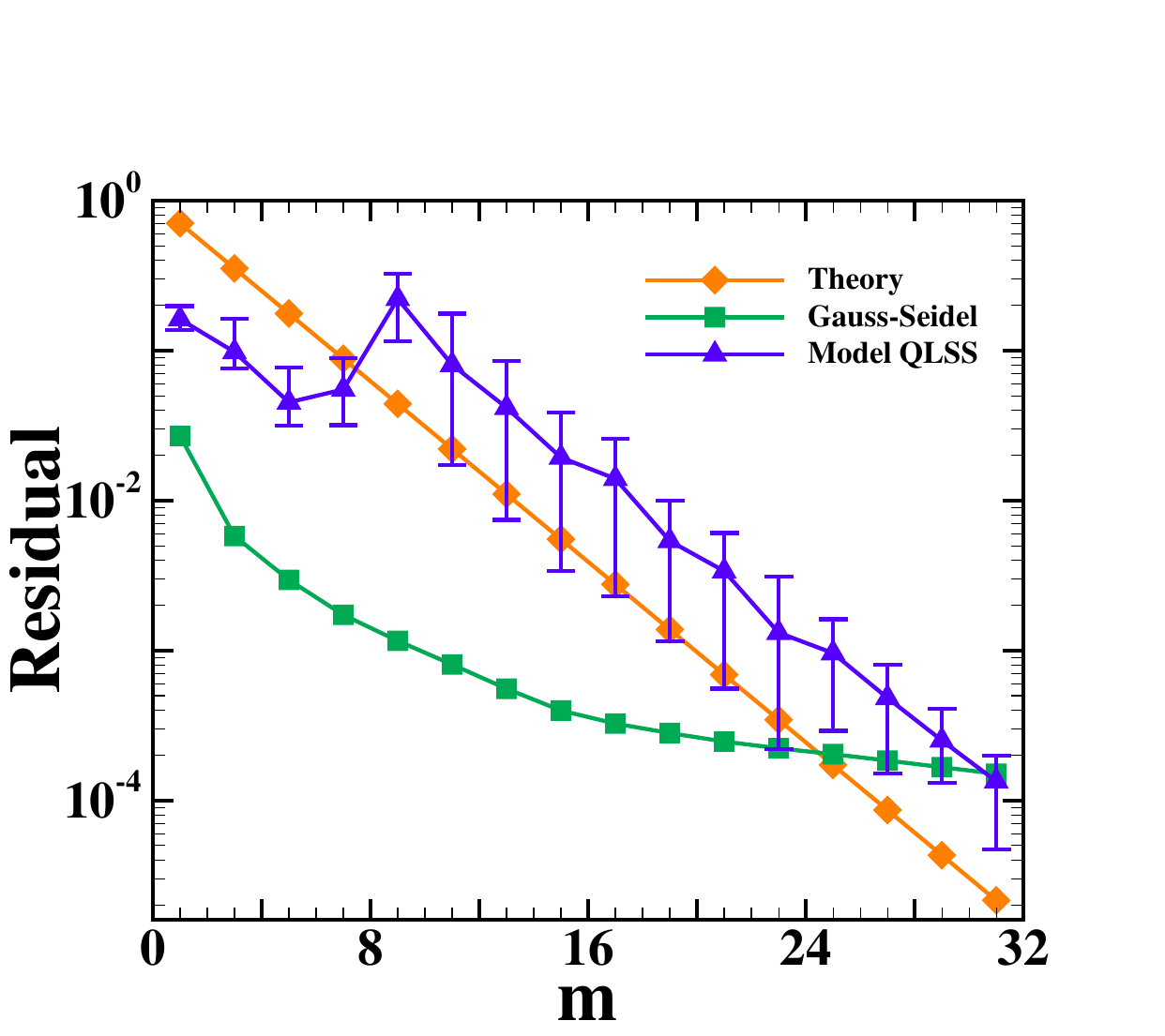}
\caption{Average error for various $m$, $m$ being either the number of qubits used for the QPE in model QLSS or the number of iterations used for Gauss-Seidel algorithm. The error bars are given by the best and worst residuals obtained by the model QLSS. 'Theory' denotes the scaling of the error bound for using $m$ qubits.}
\label{fig:RandomMatricesPerformance}
\end{subfigure}%
\caption{The model QLSS performance for different accuracies, determined by $m$ the number of qubits of the QPE. The method is tested on $10$ problems of size $N=64$ with randomly generated semi-positive definite matrices with dominant diagonals and normalized randomly generated vectors.}
\end{figure}

The proposed QLSS can be applied directly to solve the linear system of a discretized linear PDE, which can be obtained for example by applying finite difference schemes. Consider the steady two-dimensional linear Advection-Diffusion equation given by
\begin{equation}
    v\nabla u - D \nabla^2 u = f,
\end{equation}
with periodic boundary conditions, $D\in\mathbb{R}$ and $u(x,y): [0, 2\pi]^2\rightarrow \mathbb{R}$ being unknown. Further we adapt characteristics of a Taylor Green vortex, similar to the work of Over et al. \cite{Over.10.10.2024Vortex}, so the velocity is defined as 
\begin{equation}
    v(x,y) = \begin{pmatrix} \sin(x)\cos(y) \\ -\cos(x)\sin(y)\end{pmatrix}.
\end{equation}
To obtain a non trivial steady state solution of such a problem we define $f(x,y)= D \sin(x)\sin(y)$. For all simulations presented herein $D=0.25$ was chosen. Using a simple finite difference scheme one can describe this PDE as a linear system $Au = b$. The normalized solution for such a problem is shown in Figure~\ref{fig:vortexWangSol}.
To compare the results of a single QLSS solution we rescale the quantum result to the original solution. This is necessary because the amplitude encoded solution is normalized. Further, a naive measurement approach would only give positive amplitudes and could not distinguish between positive and negative signs. In practice, one must therefore employ a measurement scheme to obtain the relative signs of the solution values, see \cite{Manzano.2023RealAmpEst} for example.\\
For different problem sizes $N$ the error should, theoretically, stay constant since the error scales with the number of qubits used for the QPE, not the problem size. Any such increase would make this method useless for the use cases in CFD, which surpass these problem sizes by several orders of multitudes. As can be seen in Figure~\ref{fig:errorPlotExp1}, for a very small sample of problem sizes the error stays in the same order of magnitude, even indicating the error might decrease for increasing problem sizes. It is suspected that this is not a general trend and rather an artifact due to the small sample size. The relative error approaches the theoretical error bound and is not suspected to decrease further.\\
As can be seen in Figure \ref{fig:exp3}, increasing the number of qubits used to encode the eigenvalues increases the accuracy of the QLSS. Due to the arithmetic, each new qubit used here increases the total width by four qubits, so the numerical study of the error scaling using Qiskit simulations is heavily limited by computational resources. Up to the studied number of qubits $m=3$ (or $21$ qubits in total) one can observe that the Qiskit simulations are below the theoretical error scaling of $O(2^{-\frac{m}{2}}) =O(\varepsilon)$. One can also observe that the model QLSS is performing worse than Qiskit simulations for $m=2, 3$, but then follows the theoretical decline in error. The difference between the Qiskit simulations and the model QLSS are the result of some simplifications regarding the quantum physics with regards to normalization, not differences of the underlying algorithm.\\
A more extensive study of the error for the model QLSS was done on randomly generated problems. For this, random matrices with a dominant diagonal were generated. The linear problem was then given by $A^*Ax=A^*b$, such that the matrix $A^*A$ is semi-positive definite and hermitian. In Figure \ref{fig:ErrorBySource} the error was studied in each step. Here, both the error in approximating and also the inversion of the eigenvalues by comparison follow the theoretical bound. The inversion is done on already approximated eigenvalues and small errors in that approximation are amplified by the inversion, especially for very small eigenvalues. Looking at the total error on random problems, as can be seen in Figure~\ref{fig:RandomMatricesPerformance}, the performance of the algorithm is slightly worse than the error in the two steps alone. This observation is further supported by Figure \ref{fig:exp3}. This is the result of both error sources being amplified, as mentioned.\\
While the number of qubits can not be directly compared to the number of iterations in the classical Gauss-Seidel method, both are measures of accuracies for the respective algorithms. We can see that each additional qubit for the QPE leads to stronger improvements in the solution than additional iterations in the classical method. Further, the residuals decrease exponential in the quantum method, while the improvement per Gauss-Seidel iteration seems to decrease. For high accuracies or exact solutions this indicates an advantage of the quantum algorithm over Gauss-Seidel iterations. These results show a general applicability of a full quantum algorithm to linear PDEs. It is a good indicator for the performance of the QLSS applied to matrices occurring in the area of discretized PDEs.
\subsection{Quantum Newton Method}
\label{sim:QNM}
Combined with Newton's method the QLSS is able to solve nonlinear problems. Next, such problems are investigated. First, a simple nonlinear Poisson equation is solved in Section \ref{sim:NonLinPoisson}. To highlight the applicability to stronger nonlinearities and to show a possible future applicability for industry relevant use-cases in Section \ref{sim:NonLinProbBurgers} the Burgers equation will be solved.
\begin{figure}[tb]
\begin{minipage}[t]{0.4\linewidth}
\includegraphics[width=\linewidth]{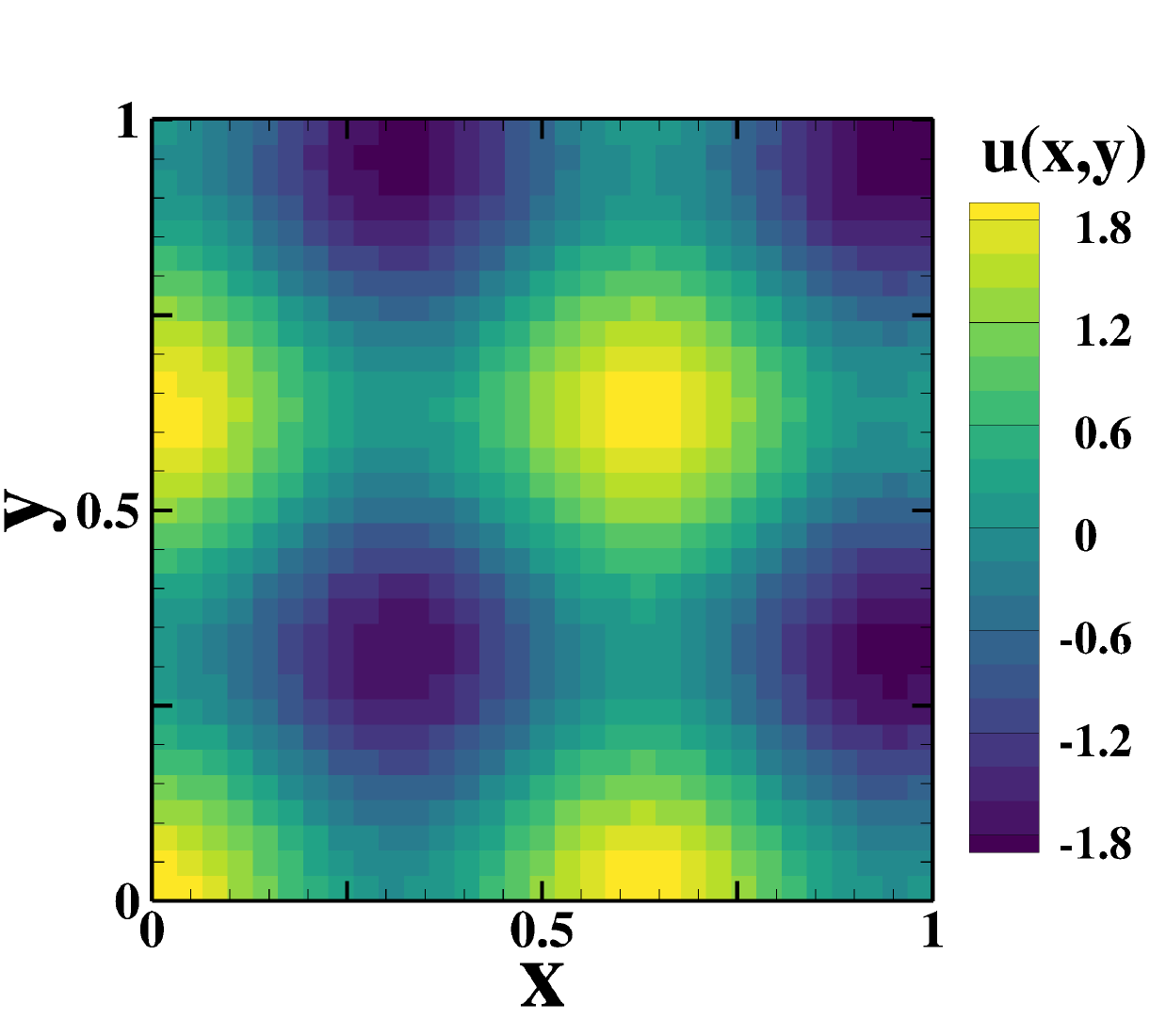}
\caption{The numerical solution of the stated nonlinear Poisson equation with $N=32$ obtained by model QLSS with $m=20$.}
\label{fig:PoissonSol}
\end{minipage}
\hfill
\begin{minipage}[t]{0.4\linewidth}
\includegraphics[width=\linewidth]{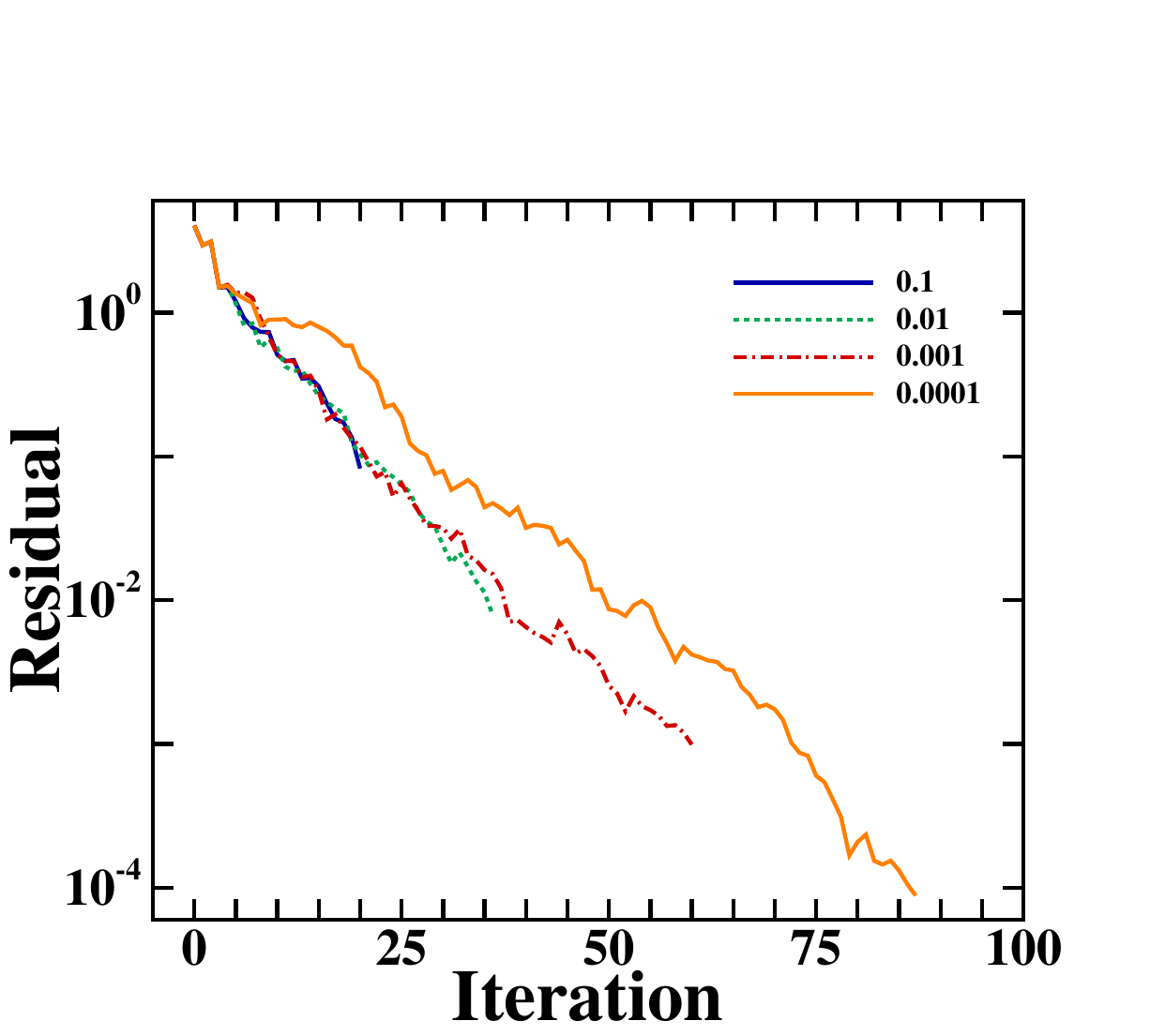}
    \caption{Convergence plot for different residual thresholds as stopping criteria of Newton's method using a Qiskit simulated QLSS with $m=2$ for the Poisson equation with $N=8$. The randomness from the various runs comes from the shot-based Qiskit simulation.}
    \label{fig:convQNMPoisson}
\end{minipage}%
\end{figure}
\subsubsection{Poisson Equation}
\label{sim:NonLinPoisson}
\begin{figure}
\begin{minipage}[t]{0.4\linewidth}
\includegraphics[width=\linewidth]{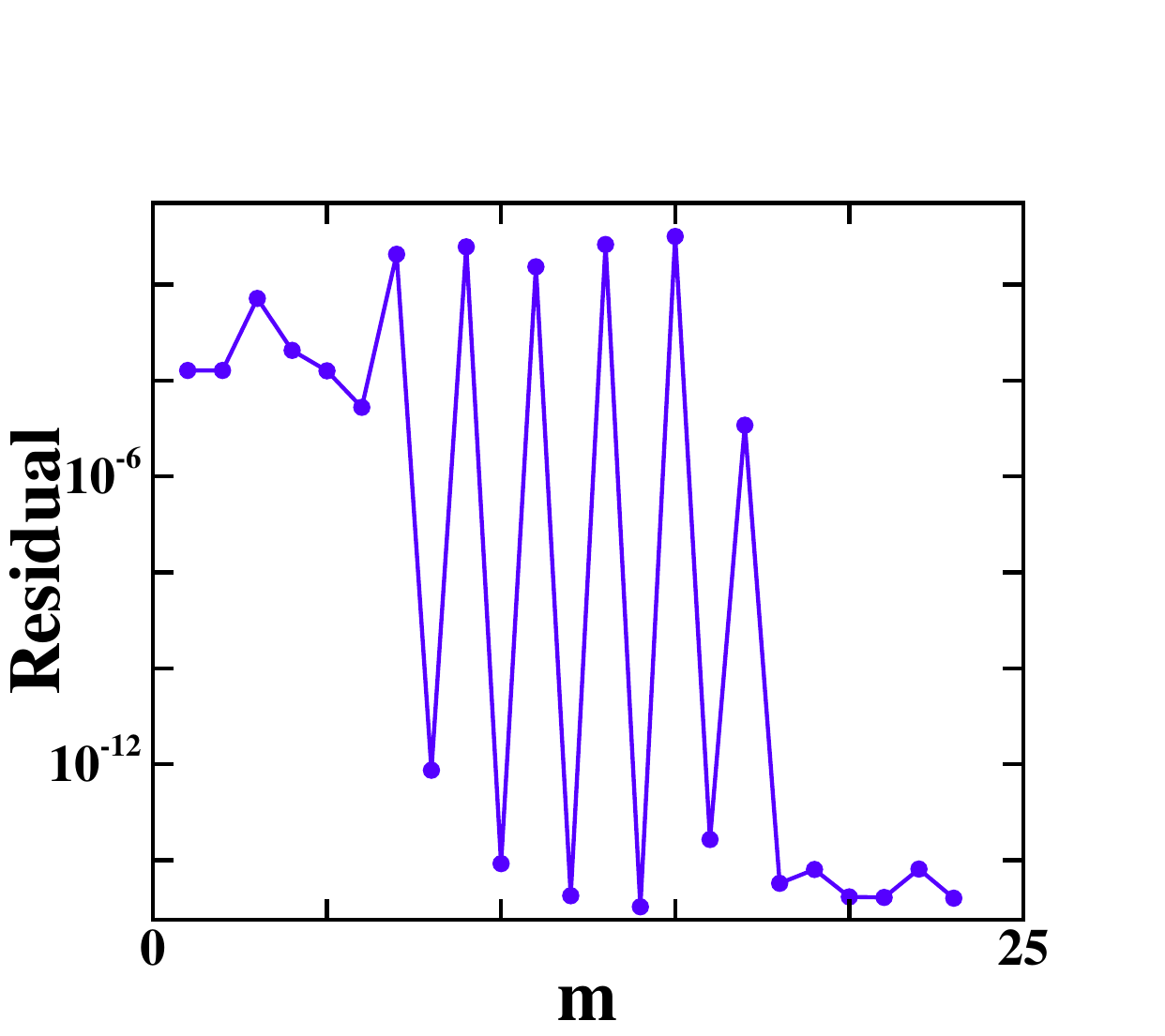}
    \caption{Residual reached for the QNM using model QLSS within $50$ iterations for a Poisson equation with $N=8$.}
    \label{fig:Poisson50It}
\end{minipage}
\hfill
\begin{minipage}[t]{0.4\linewidth}
    \includegraphics[width=\linewidth]{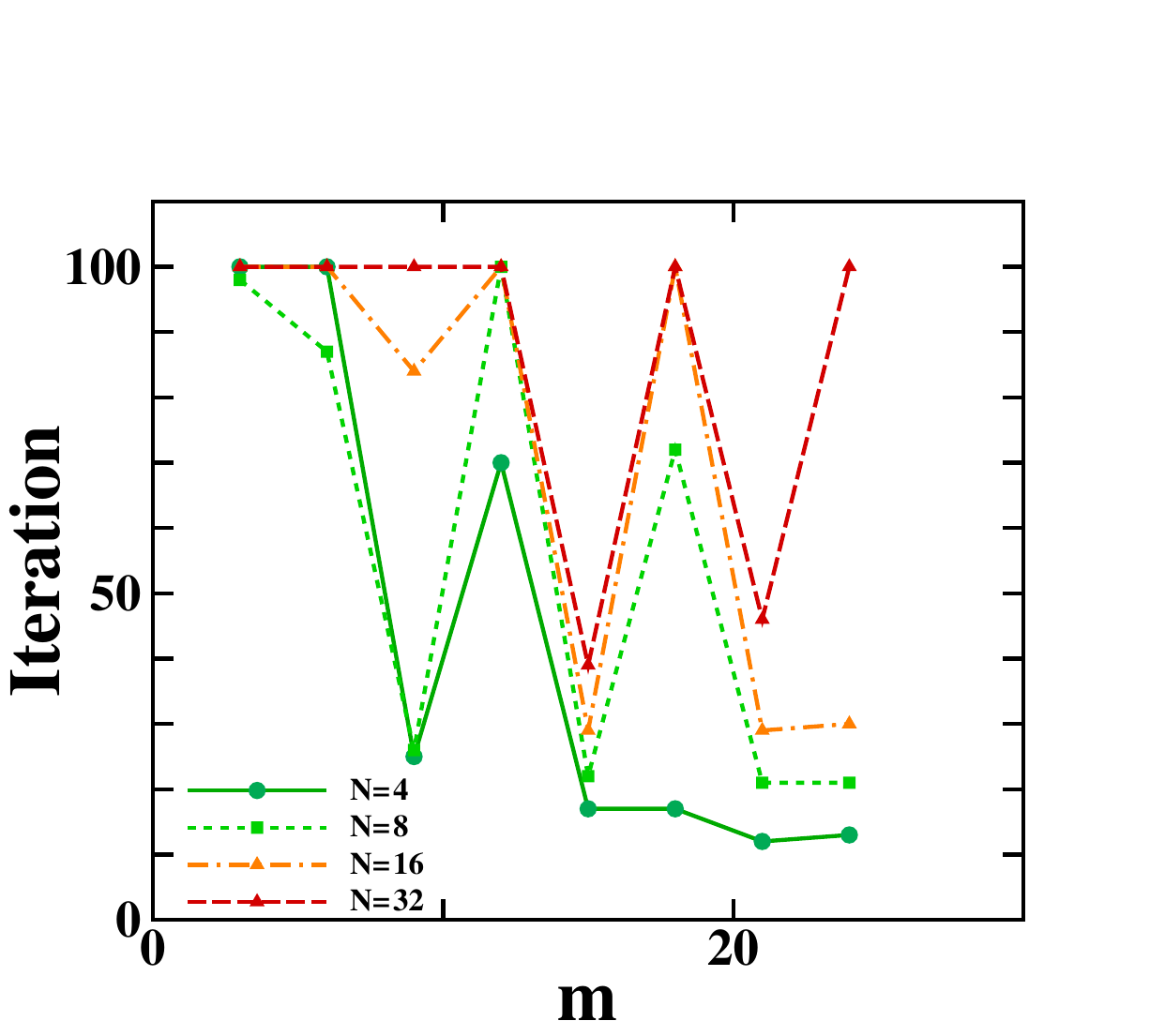}
    \caption{Number of iterations necessary for the QNM to reach a residual below $10^{-6}$ for various choices of $m$ and $N$ for the Poisson equation. The number of iterations is limited to $100$.}
    \label{fig:PoissonHeatmap}
\end{minipage}
\end{figure}

The Poisson equation is a widely used model problem to compare different methods \cite{Trottenberg.2007Multigrid} and is given by
\begin{align*}
    -\Delta u&=f(x)\quad,x\in\Omega\subset\mathbb{R}^d \\ 
        u &= f_b(x)\quad,x\in \partial\Omega,
\end{align*}
where $d$ is usually $2$ or $3$. In our two-dimensional example, i.e. $d=2$, we consider a nonlinear function $f(u,x,y)=C^2\cos(Cx)\cos(Cy) + u(x,y)^2$ with $C=10$. We further define $\Omega=[0, 1]^2$ and $f_b(x,y)=\cos(Cx)\cos(Cy)$. A solution obtained by QNM using model QLSS can be seen in Figure \ref{fig:PoissonSol}.\\
Consider a two-dimensional discretized Poisson equation with $8\times 8$ grid points, i.e. $N=8$, and functions as above. A Qiskit simulation is ran for a QNM with $m=2$, so the error in estimating the eigenvalues is of order $2^{-1}=0.5$. As can be seen in Figure \ref{fig:convQNMPoisson}, one can still obtain accurate solutions at the cost of running more Newton iterations. One can even obtain a solution multiple orders more accurate than $m$, the number of QPE qubits, would suggest. The inaccurate representation of the eigenvalues lead to some iteration steps increasing the residual, i. e. worsening the solution.\\ 
\begin{figure}[htb]
\begin{subfigure}[t]{0.45\textwidth}
\centering
\includegraphics[width=\linewidth]{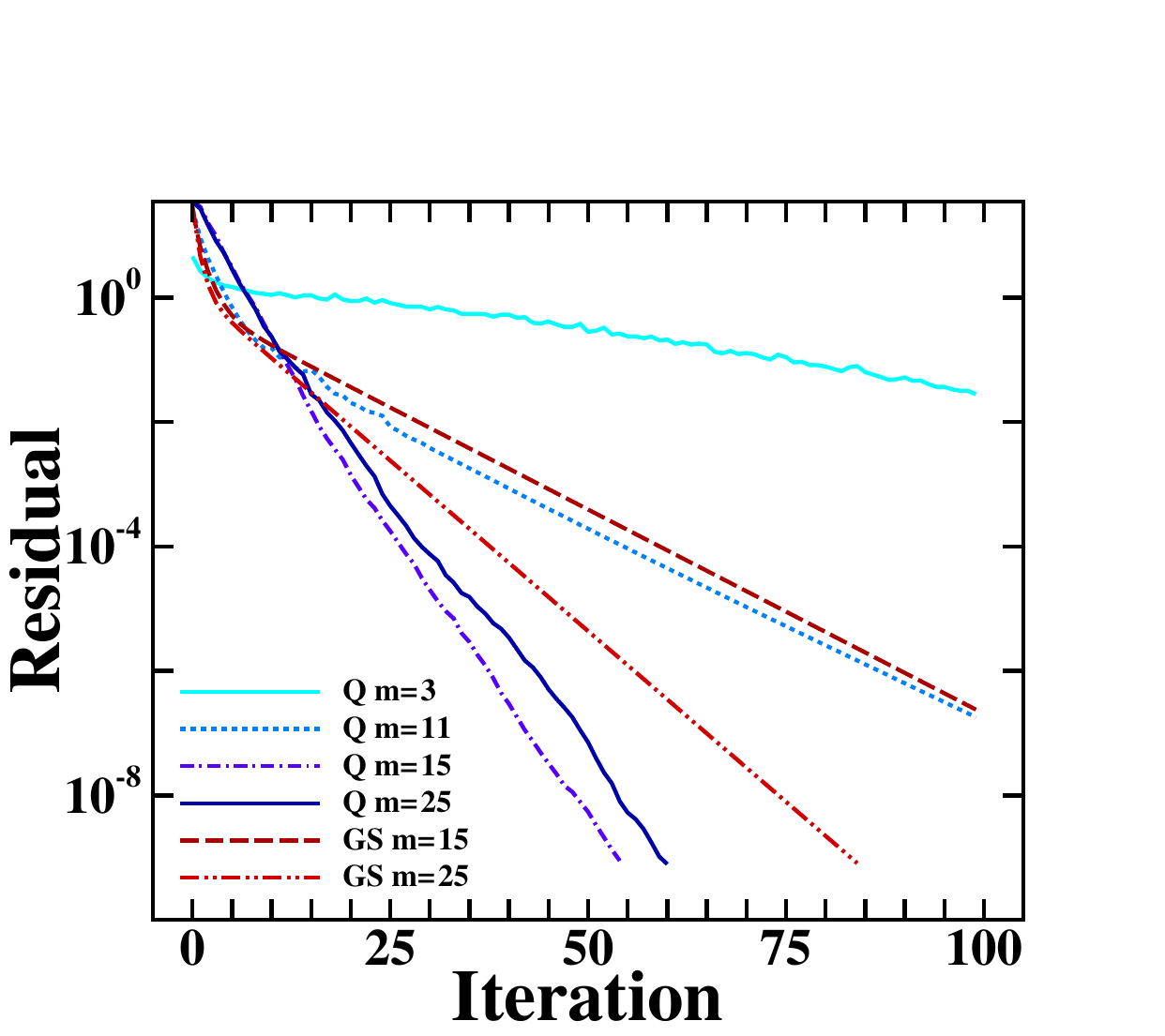}
\caption{Varying $m$ qubits for the QPE ('Q')/ number of Gauss-Seidel iterations ('GS') on a problem with size $N=32$}
\label{fig:PoissonMQNM_mStudy}
\end{subfigure}%
\hfill
\begin{subfigure}[t]{0.45\textwidth}
\centering
\includegraphics[width=\linewidth]{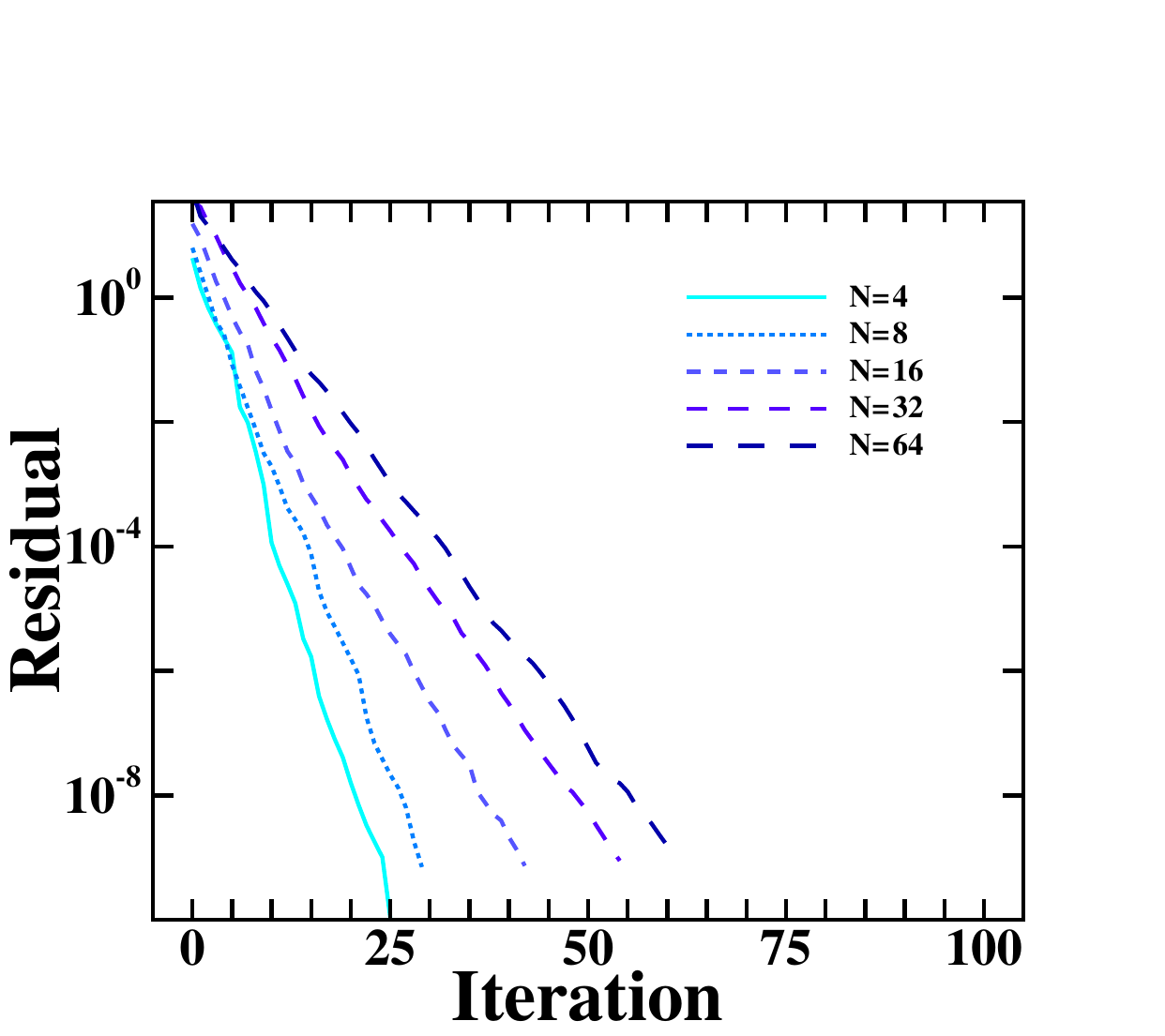}
\caption{Varying problem sizes $N$ with $m=20$ qubits for the QPE.}
\label{fig:PoissonMQNMSizes}
\end{subfigure}%
\caption{Convergence plots for the QNM using model QLSS for the nonlinear Poisson equation. }
\end{figure}
Using the previously discussed model QLSS, we can study the presented method up to higher accuracies and obtain insights into its scaling behavior.
In Figure \ref{fig:PoissonMQNM_mStudy} we compare the convergence for various choices of accuracy, determined by the number of qubits $m$ bounding the error by $2^{-\frac{m}{2}}$. The figure displays that already with $m=15$, i.e. an error bound of $\approx0.006$, convergence in less than $60$ iterations is obtained. Increasing $m$ further does not yield much improvement, as $m=20$ does not improve the performance. Further, the classical method is depicted for either $15$ or $25$ Gauss-Seidel iterations in each Newton iteration. Again, while comparing number of qubits used for the QPE and number of iterations of the Gauss-Seidel method is not a fair comparison, in Figure~\ref{fig:PoissonMQNM_mStudy} one can see that the decrease of the residual the QNM is able to achieve is faster than for Newton's method using Gauss-Seidel iterations. Considering the linear system the QLSS solves is twice the size of the Gauss-Seidel method, due to it not being hermitian, this is a surprising observation. This indicates an advantage of the presented method, whenever the solution to the linear system should be solved to a high degree of accuracy, but since an additional Gauss-Seidel iteration is, in general, easier to realize than accessing more qubits this should be taken with caution.
Instead of increasing $m$, in Figure \ref{fig:PoissonMQNMSizes} we fix $m=20$ and study the performance for various problem sizes. Consider the problem sizes from $N=4$ and $N=64$, a $16$ times increase. The method obtains the same residual in around $60$ instead of $25$ iterations, merely a $2.4$ times increase. This offers an improvement over time-stepping schemes, whose iterations scale linearly with $N$.\\
\begin{figure}[tb]
\begin{subfigure}{0.3\textwidth}
\centering
\includegraphics[width=\linewidth]{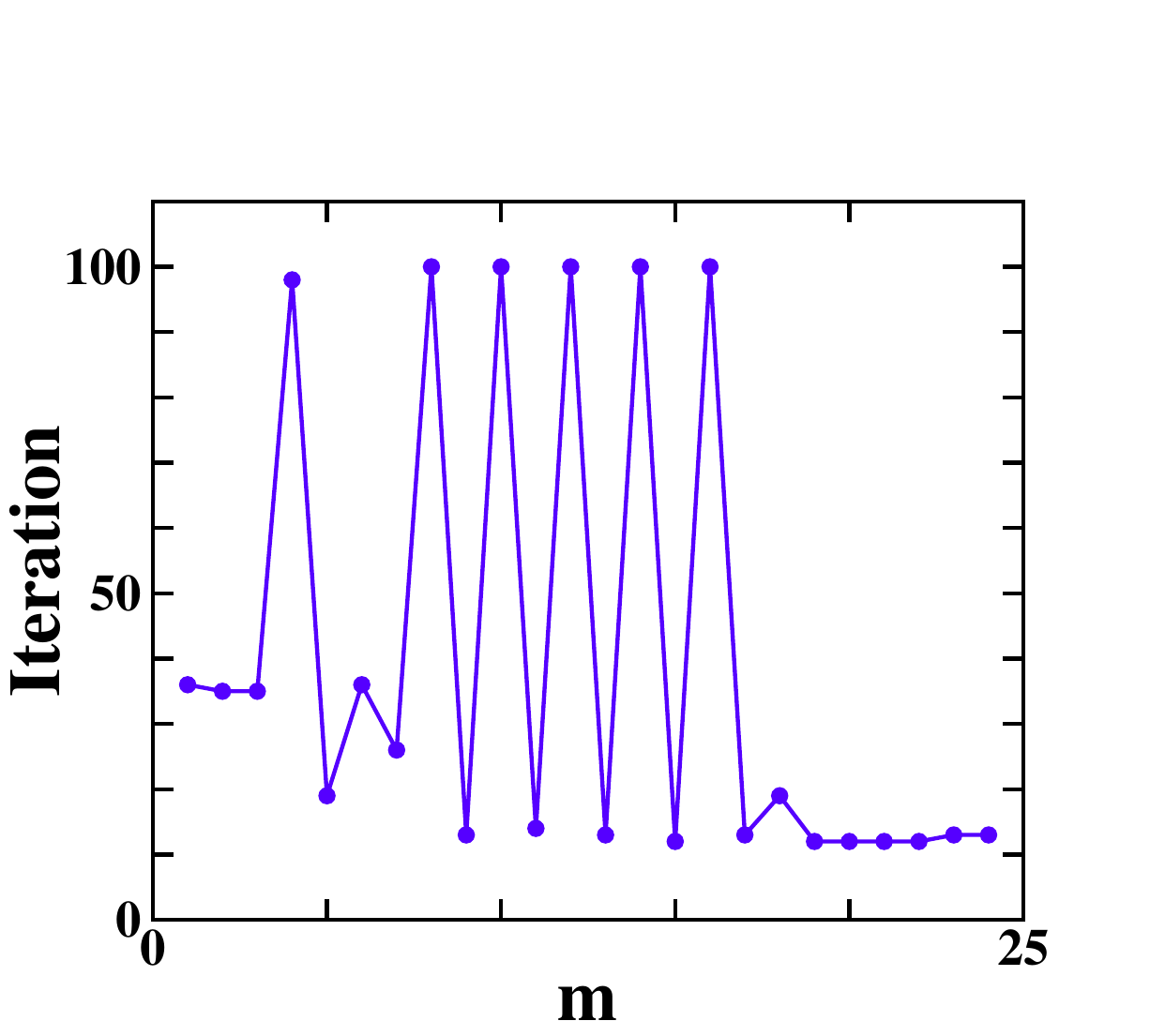}
\caption{Iterations until residual below $10^{-3}$ is obtained.}
\label{fig:PoissonMe-3}
\end{subfigure}%
\hfill
\begin{subfigure}{0.3\textwidth}
\centering
\includegraphics[width=\linewidth]{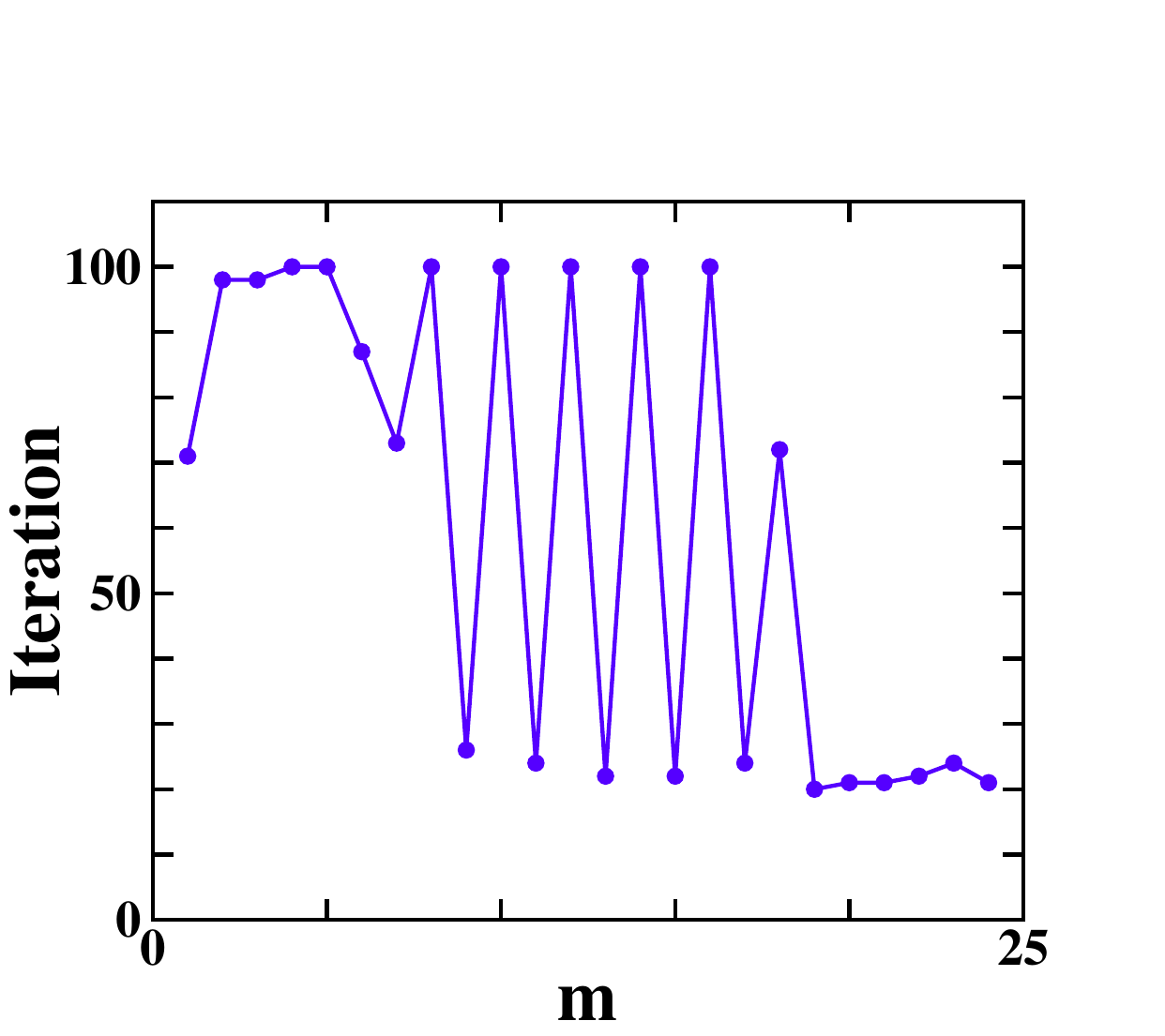}
\caption{Iterations until residual below $10^{-6}$ is obtained.}
\label{fig:PoissonMe-6}
\end{subfigure}%
\hfill
\begin{subfigure}{0.3\textwidth}
\centering
\includegraphics[width=\linewidth]{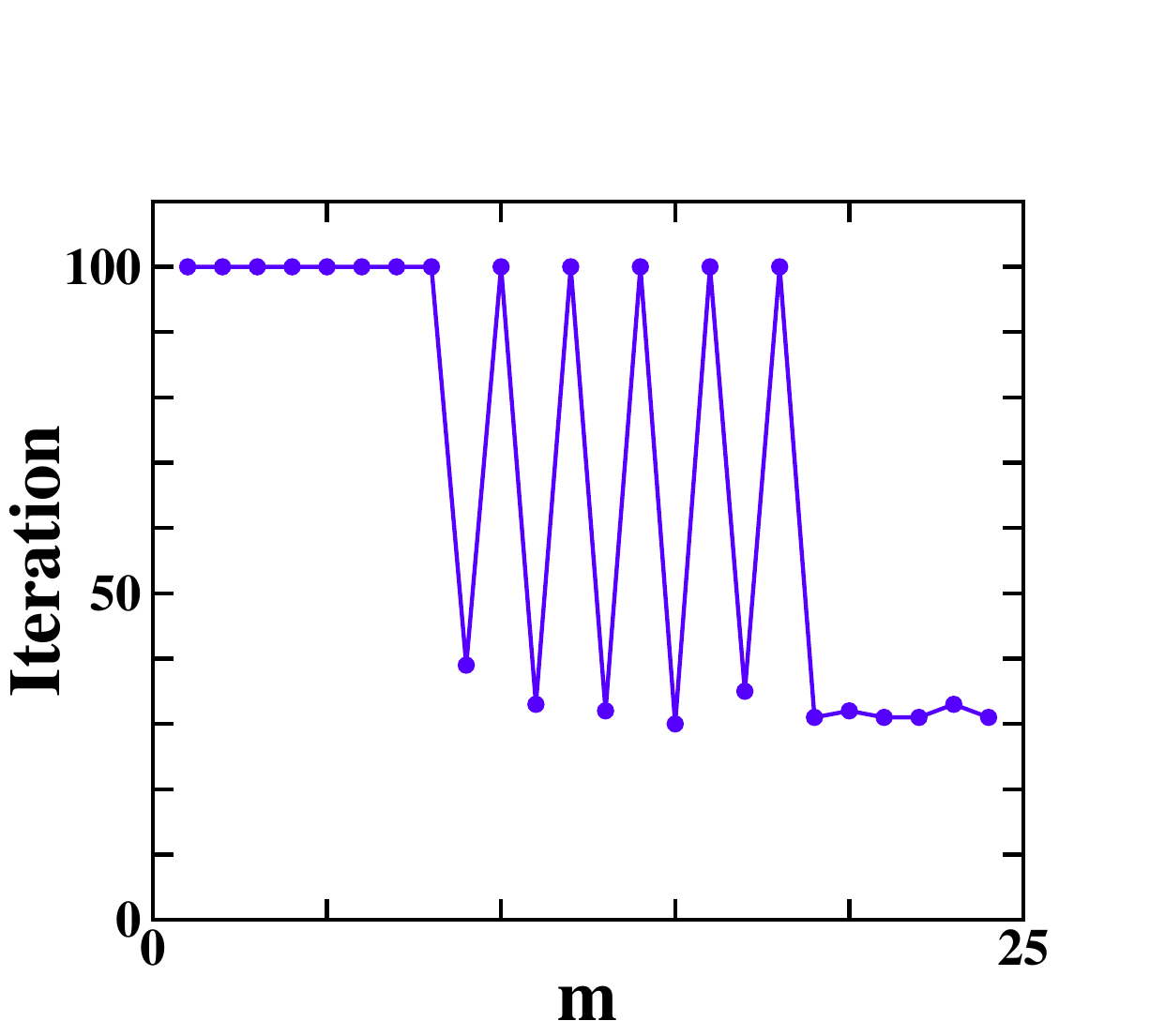}
\caption{Iterations until residual below $10^{-9}$ is obtained.}
\label{fig:PoissonMe-9}
\end{subfigure}
\caption{Number of iterations needed for the QNM using model QLSS reaching a residual below various thresholds for a Poisson equation with $N=8$. The QNM is stopped after $100$ iterations should the residual not be reached.}
\label{fig:PoissonMe-all}
\end{figure} 
Measuring the number of iterations, necessary to reach residuals below various thresholds for various choices of $m$, one obtains Figures \ref{fig:PoissonMe-3}, \subref{fig:PoissonMe-6} and \subref{fig:PoissonMe-9}. In these figures three distinct regions of behavior can be observed. For a very low number of qubits, $m=1$ to $m=9$, the method does reach the residual of order $10^{-3}$, but for the lower thresholds the maximum number of $100$ iterations is reached beforehand. Afterwards, for $m=10$ to $m=19$ an unstable behavior is depicted. Either the method achieves residuals below $10^{-9}$ in less than $50$ iterations, or $100$ iterations pass without the QNM obtaining a residual of $10^{-3}$. Increasing $m$ further, does not decrease the number of iterations necessary for a full convergence, but importantly, removes the unstable behavior completely. For all $m\geq 20$, i.e. an error bound below $\approx 10^{-3}$, the lowest residual is reached, but notably increasing $m$ does not seem to have a strong effect on the number of iterations necessary.\\
Now in Figure \ref{fig:Poisson50It}, fix the number of iterations to $50$ and measure the size of the residuals that can be obtained at the end. The same three regions mentioned before can be observed. Notably, for the cases, whose residual reached below $10^{-9}$, full convergence, i.e. residuals of machine precision where achieved. For the converged $m$ in the unstable regions, and $m\geq 19$ we observe that increasing $m$ keeps slightly decreasing the final residual.
Finally, studying various problem sizes and the QNM behavior for various $m$ in Figure \ref{fig:PoissonHeatmap}, we observe a shift of the three regions. For example the unstable region for $N=4$ becomes stable for around $m=15$, but for $N=16$ this is only observed for $m>20$. The number of iterations is generally increased for larger $N$, as also observed in the previous Figure \ref{fig:PoissonMQNMSizes}. Therefore, an increase of $m$, and therefore the accuracy of the QLSS, does offer an improvement in stability, but can not counteract the increase in problem size completely. 
\subsubsection{Burgers Equation}
\label{sim:NonLinProbBurgers}
\begin{figure}
    \begin{minipage}[t]{0.4\linewidth}
        \includegraphics[width=\linewidth]{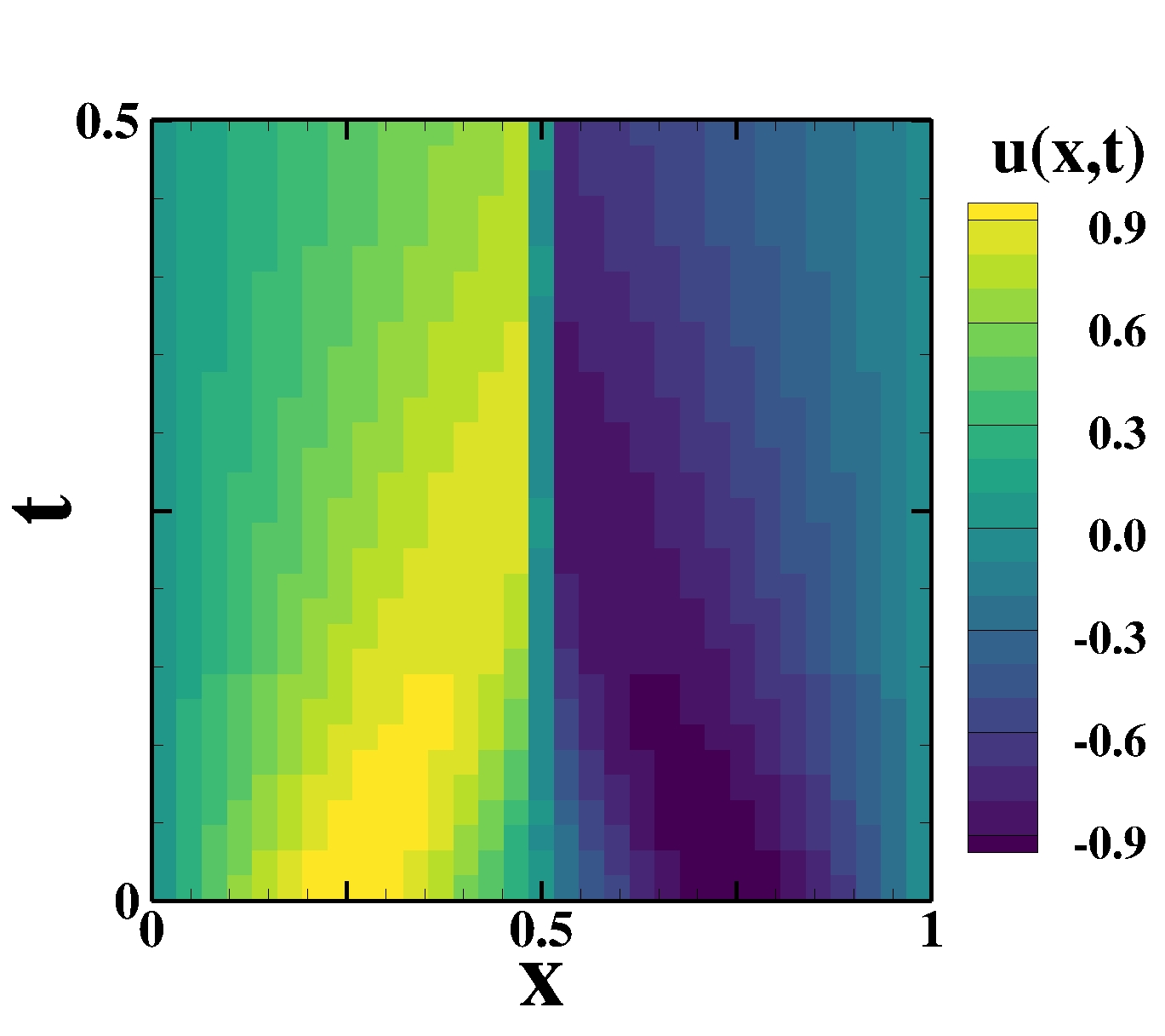}
        \caption{The numerical solution of the stated nonlinear Burgers equation on a $32\times32$ grid obtained by model QLSS with $m=20$.}
        \label{fig:BurgersSol}
    \end{minipage}
    \hfill
    \begin{minipage}[t]{0.4\linewidth}
        \centering
        \includegraphics[width=\linewidth]{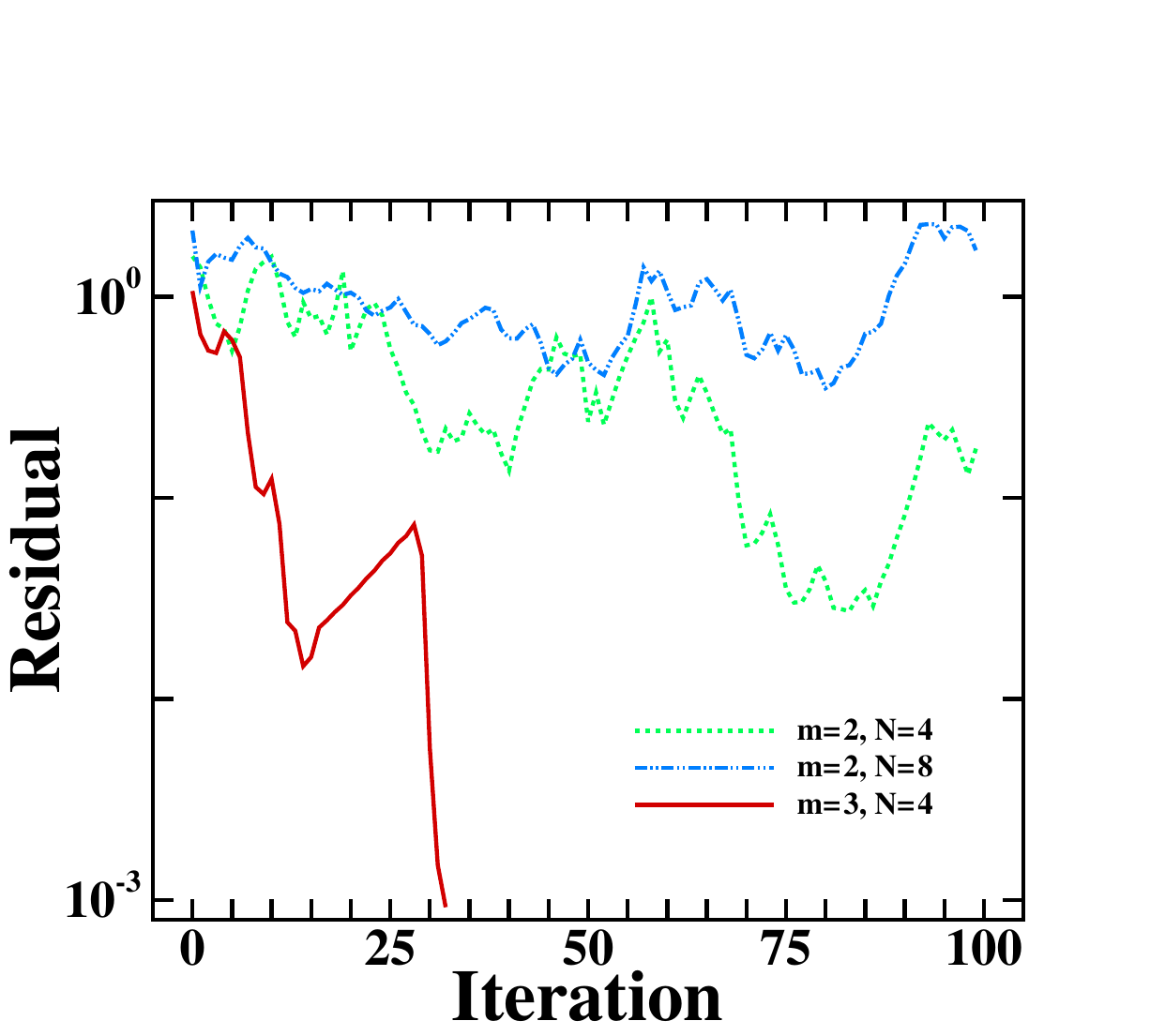}
        \caption{QNM Convergence for a $m=2$ or $3$ obtained by Qiskit simulations. QNM stopping at residual being smaller than $10^{-3}$. The Problem being solved is the Burgers equation with $N=4$ or $N=8$.}
        \label{fig:ConvBurgersEqQNM}
    \end{minipage}
\end{figure}

\begin{figure}[tb]
\begin{subfigure}[t]{0.45\textwidth}
\centering
\includegraphics[width=\linewidth]{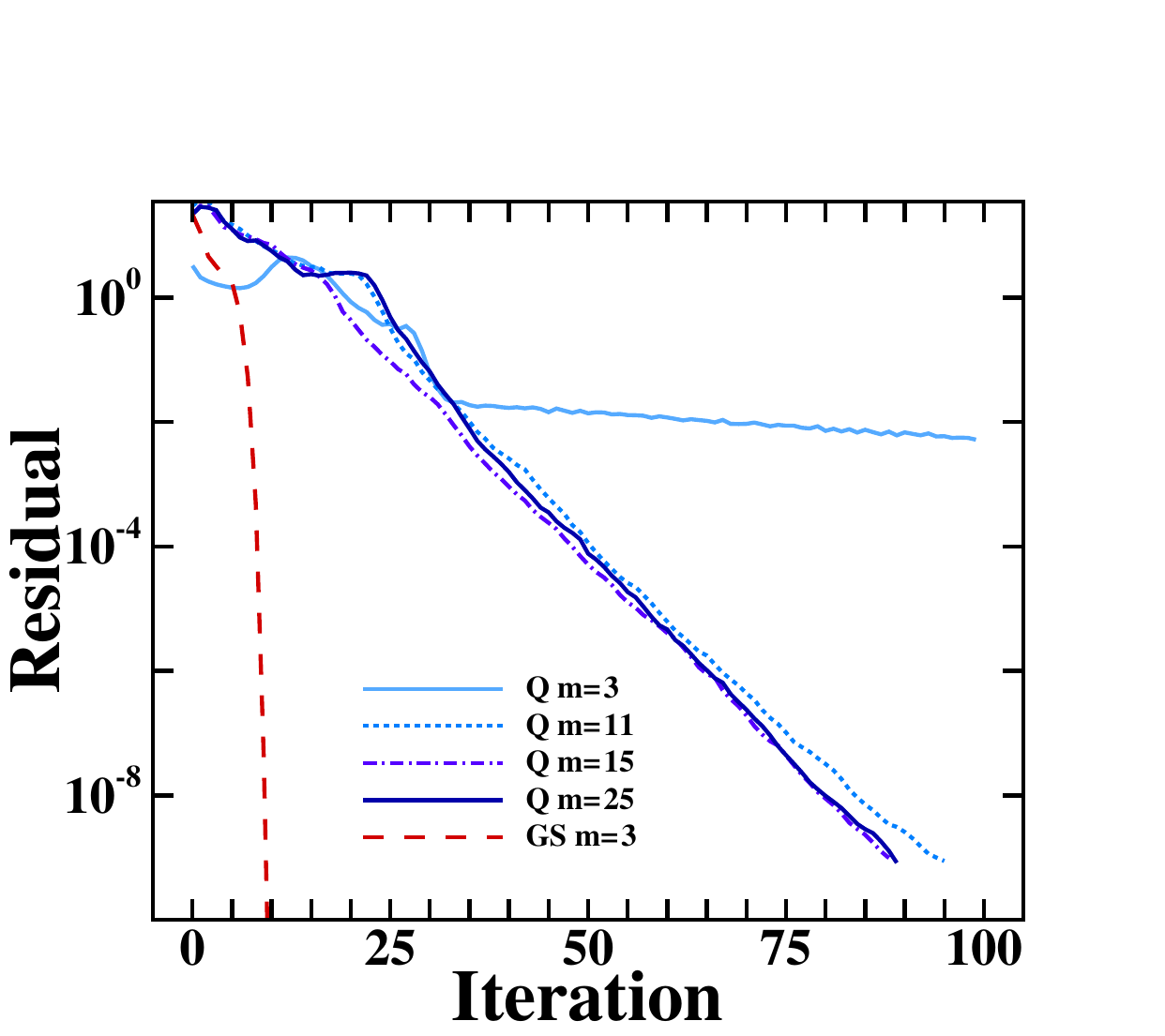}
\caption{{Varying $m$ qubits for the QPE ('Q')/ number of Gauss-Seidel iterations ('GS') on a problem with size $N=32$}}
\label{fig:BurgersMQNM_mStudy}
\end{subfigure}%
\hfill
\begin{subfigure}[t]{0.45\textwidth}
\centering
\includegraphics[width=\linewidth]{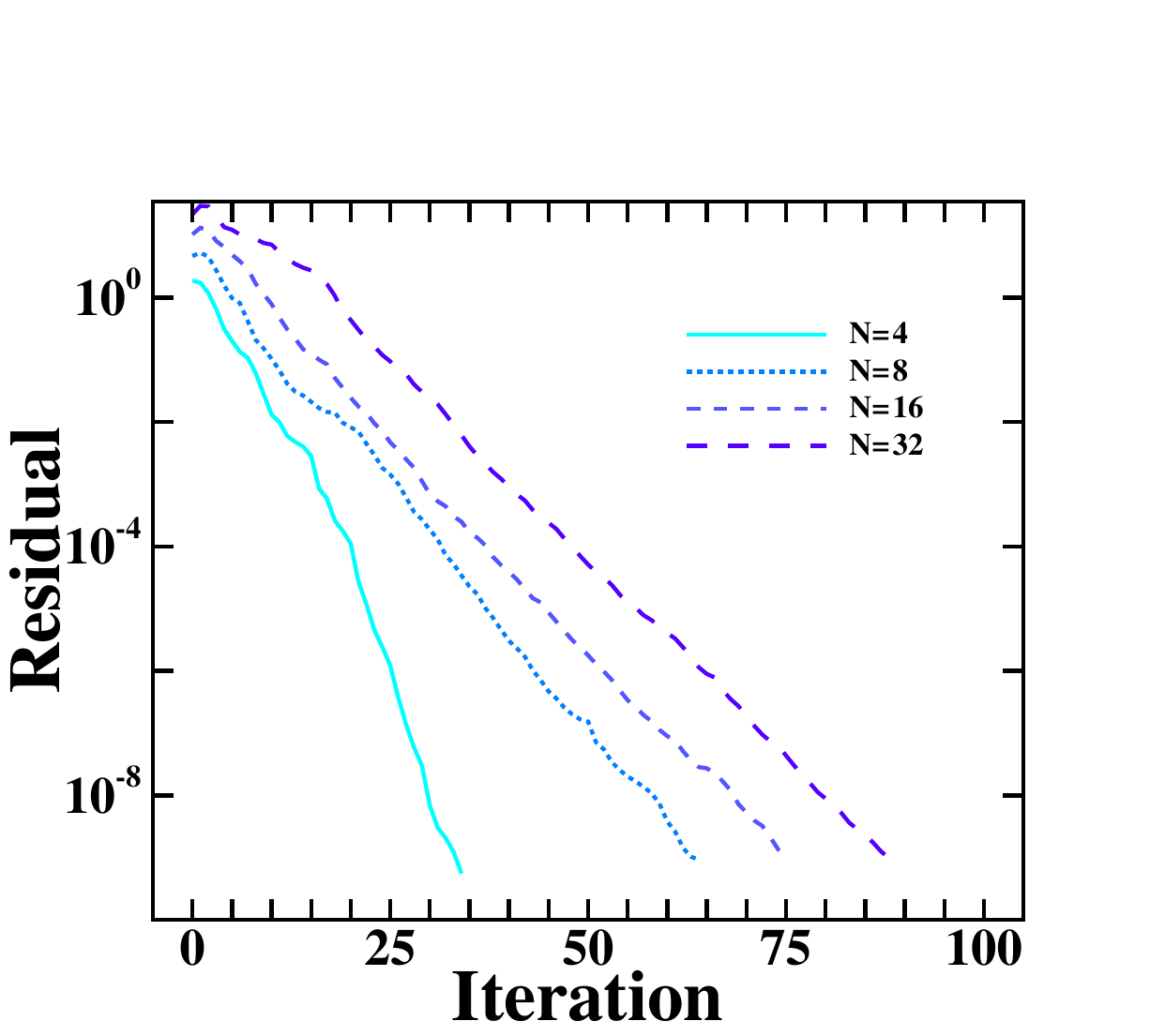}
\caption{Varying problem sizes $N$ with $m=20$ qubits for the QPE.}
\label{fig:BurgersMQNMSizes}
\end{subfigure}%
\caption{Convergence plots for the QNM using model QLSS for the Burgers equation.}
\end{figure}

\begin{figure}[tb]
    \begin{minipage}[t]{0.4\linewidth}
    \centering
    \includegraphics[width=\linewidth]{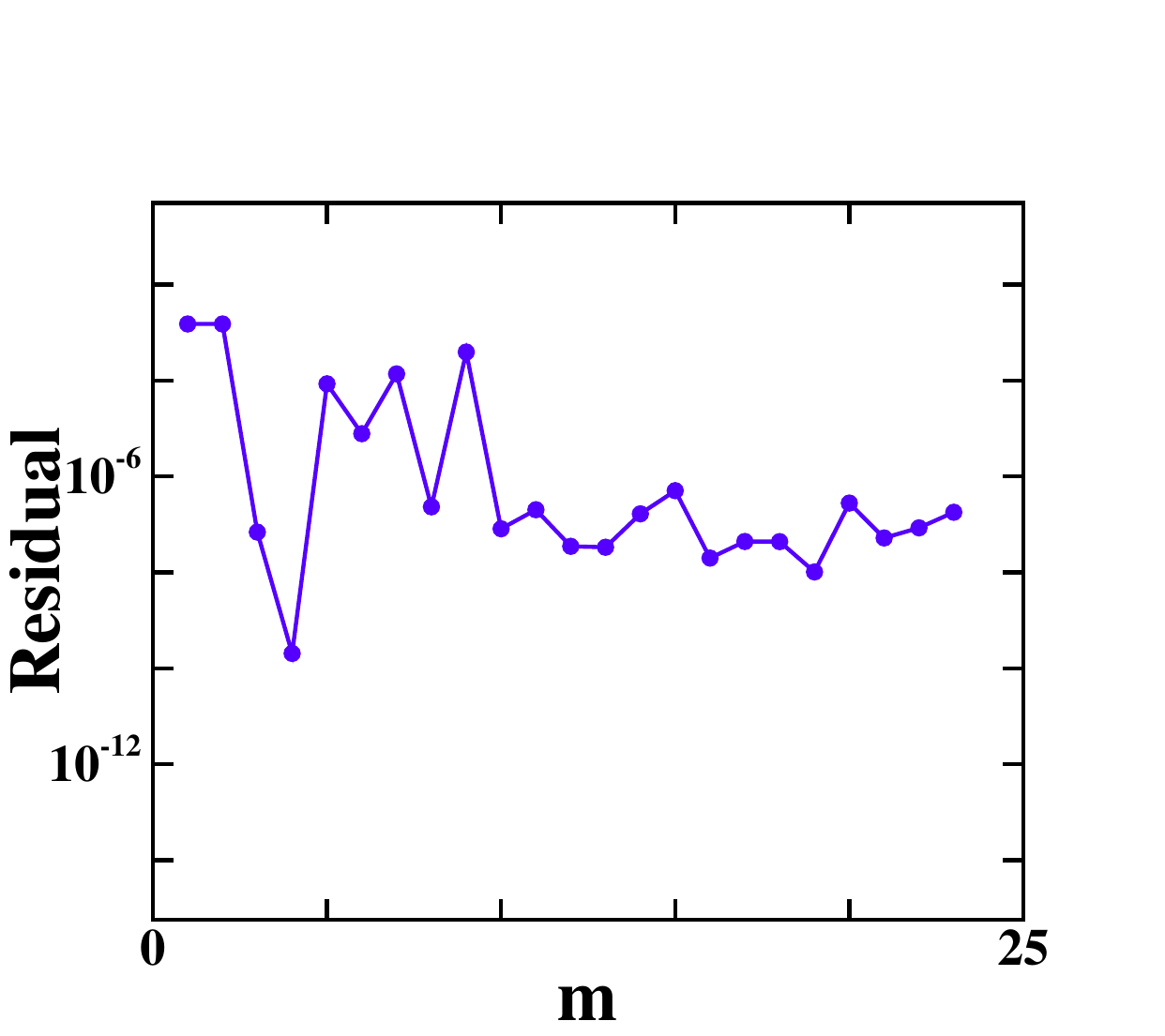}
    \caption{Residual reached for the QNM using model QLSS within 50 iterations
    for a Burgers equation with $N=8$.}
    \label{fig:Burgers50It}
    \end{minipage}
    \hfill
    \begin{minipage}[t]{0.4\linewidth}
        \includegraphics[width=\linewidth]{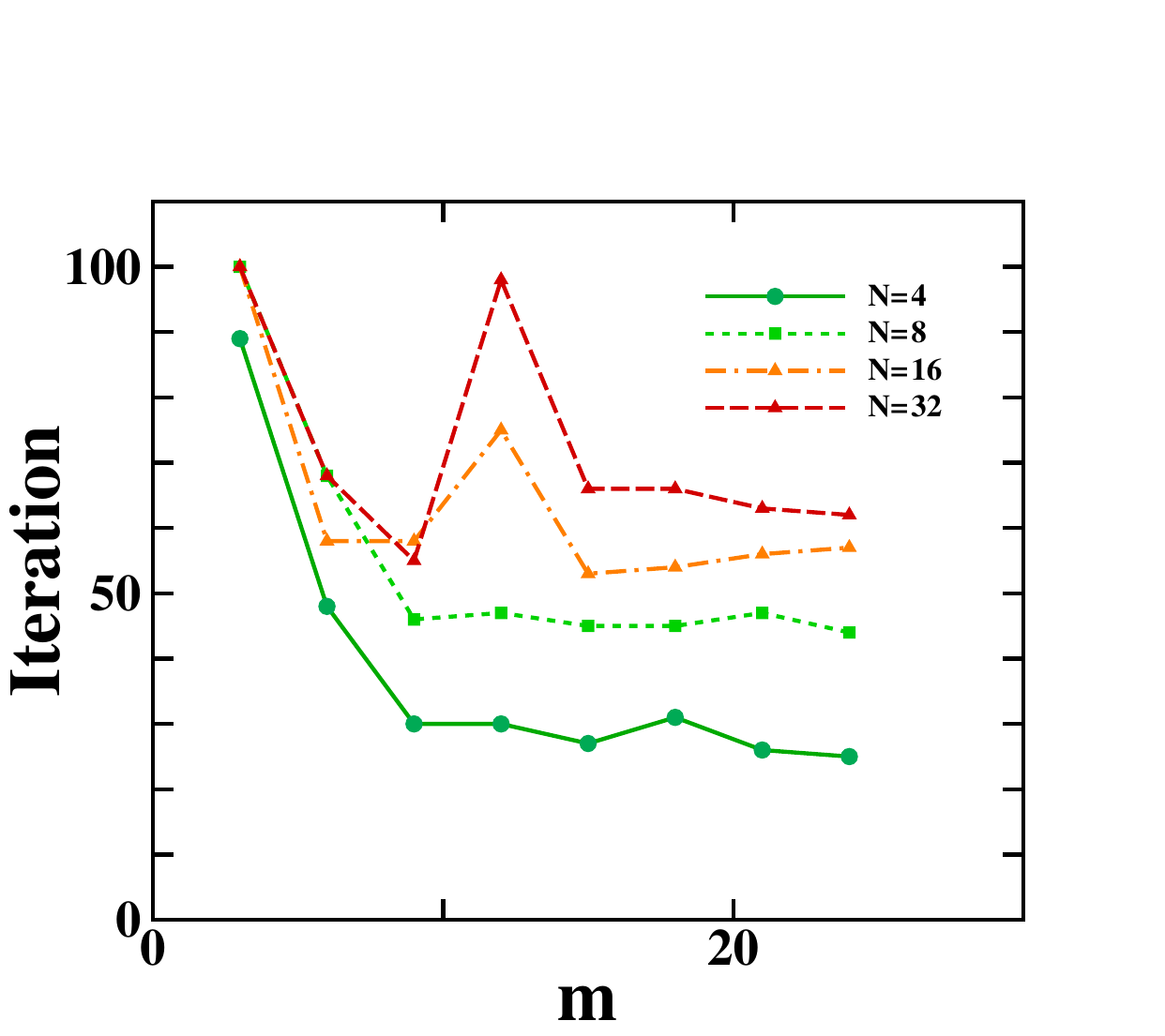}
        \caption{Number of iterations necessary for the QNM to reach a residual below $10^{-6}$ for various choices of $m$ and $N$ for the Burgers equation. The number of iterations is limited to $100$.}
        \label{fig:BurgersHeatmap}
    \end{minipage}
\end{figure}

The Burgers equation is a widely known example for a nonlinear PDE. It was first introduced by Bateman \cite{BATEMAN.1915_OGBurgers} and has been used in many different fields. The equation of the inviscid Burgers equation is given by 
\begin{equation}
    \frac{\partial u}{\partial t} + u \frac{\partial u}{\partial x} = 0.
\end{equation}
Consider the initial condition of $u(x, 0)=\sin{2\pi x} $, for $x\in [0,1]$ and further consider the boundary conditions $u(0, t) = u(1, t) = 0$ for all $t\in[0, 0.5]$. We discretize both in space and time with an equal amount of grid points, meaning the temporal discretization is twice as fine as the spatial. Moreover, we use a finite difference scheme with a simple first-order upwinding scheme. This setup, as can be seen in Figure~\ref{fig:BurgersSol}, will produce a shock, a phenomena which is hard to reproduce, especially for global linearization methods due to its discontinuous nature.\\
Performing Qiskit simulations the problem did not converge using only $m=2$ qubits for the QPE and so at least $m=3$ qubits were required, as can be seen in Figure \ref{fig:ConvBurgersEqQNM}. The convergence is, as expected, worse than for the Poisson equation (compare Figure \ref{fig:convQNMPoisson}) and it even diverges for many iterations. The use of more iterations improves the residual orders of magnitude below the QLSS error bounds. Increasing the size of the problem from $N=4$ to $N= 8$, does increase the difficulties of convergence. For $m=2$ one can observe that the smaller problem results in lower residuals but does not achieve convergence yet. Overall this effect appears to be less significant than the effect of increasing $m$, as the residual for $m=3$ decreases a lot faster.\\
QNM using model QLSS was also applied to solve the Burgers equation, offering more insights into the performance of the method.
In Figure \ref{fig:BurgersMQNM_mStudy}, multiple choices of $m$, as both number of qubits and number of Gauss-Seidel iterations, are deployed and the convergence plots are studied. The hybrid method converges nicely for $m\geq11$ and has a similar performance for $m=3$ as for the Poisson equation. Already with $m=11$, in other words with an error bound of $2^{-m/2}\approx 0.022$, further qubits or a higher accuracy in the QLSS seem to not improve the QNM. Comparing QNM to Newton's method with Gauss-Seidel iterations, see 'GS $m=3$' in Figure \ref{fig:BurgersMQNM_mStudy}, the presented hybrid method is far worse and the classical scheme converges nearly immediately even with just a few Gauss-Seidel iterations. It is suspected that this is due to the inversion of the eigenvalues, which is numerically more unstable compared to the multiplications in the Gauss-Seidel method, especially for very small eigenvalues. Nevertheless, the QNM does converge, showing the application of quantum computation to highly nonlinear PDEs.
Studying QNM for various problem sizes we obtain the convergence plots in Figure \ref{fig:BurgersMQNMSizes}. Here, increasing $N$ from $4$ to $32$, a $8$ times increase, results in an increase in iterations necessary to obtain a residual of $10^{-9}$. The necessary iterations increase from $35$ to $90$, roughly a $2.6$ times increase. Again, this is an improvement over any time-stepping schemes, where the number of iterations would coincide with the number of time steps, in our case $N$. The number of iterations scales worse than for the Poisson equation, which due to the nonlinearity was expected.\\
\begin{figure}[tb]
\begin{subfigure}{0.3\textwidth}
\centering
\includegraphics[width=\linewidth]{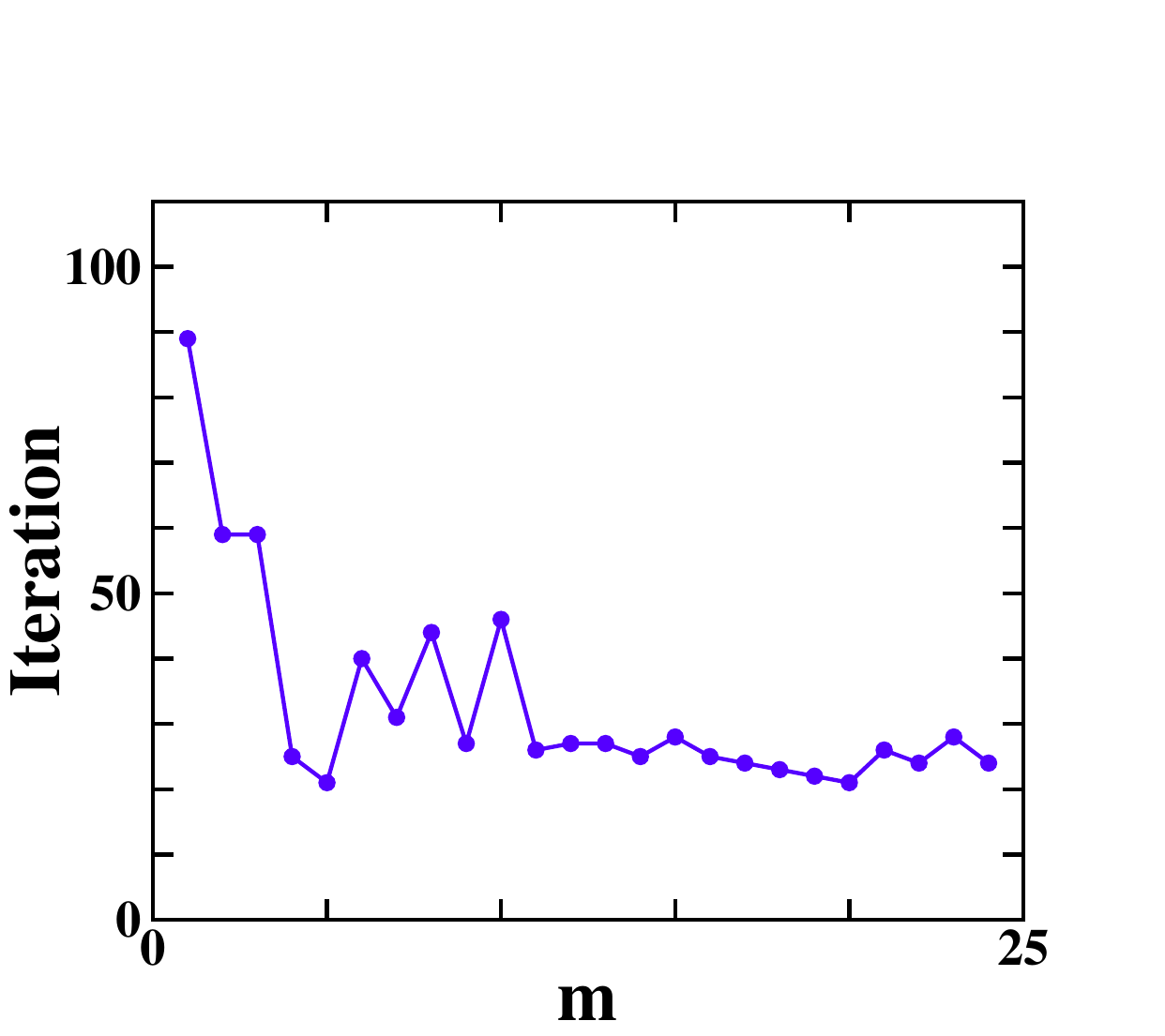}
\caption{Iterations until residual below $10^{-3}$ is obtained.}
\label{fig:BurgersMe-3}
\end{subfigure}%
\hfill
\begin{subfigure}{0.3\textwidth}
\centering
\includegraphics[width=\linewidth]{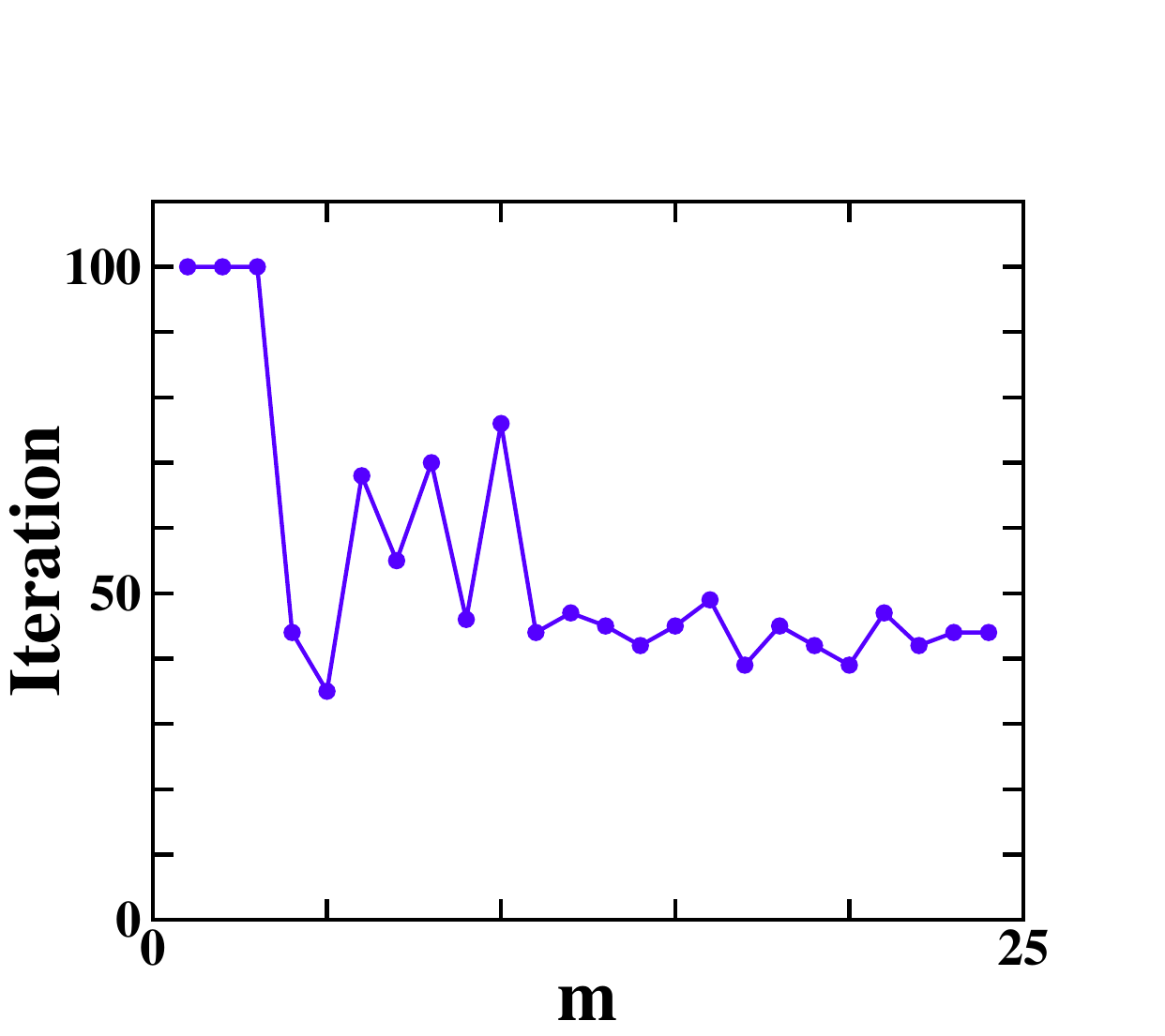}
\caption{Iterations until residual below $10^{-6}$ is obtained.}
\label{fig:BurgersMe-6}
\end{subfigure}%
\hfill
\begin{subfigure}{0.3\textwidth}
\centering
\includegraphics[width=\linewidth]{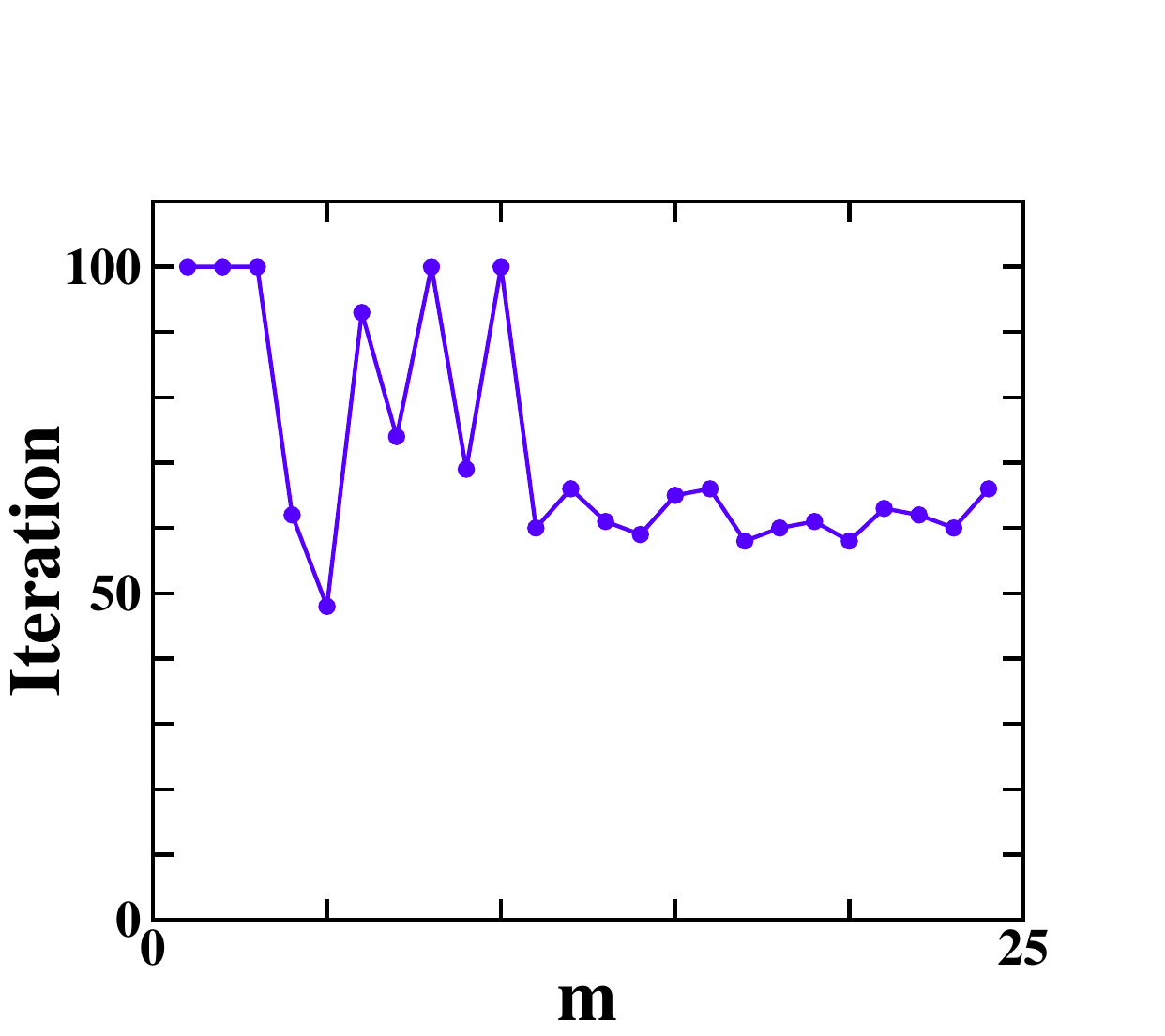}
\caption{Iterations until residual below $10^{-9}$ is obtained.}
\label{fig:BurgersMe-9}
\end{subfigure}
\caption{Number of iterations needed for the QNM using model QLSS reaching a residual below various thresholds for a inviscid Burgers equation with $N=8$. The QNM is stopped after $100$ iterations should the residual not be reached.}
\end{figure}
Denoting the number of iterations necessary for certain levels of residuals in Figures~\ref{fig:BurgersMe-3}, \subref{fig:BurgersMe-6} and \subref{fig:BurgersMe-9} a contrary observation is made. Here, the QNM performs more stable as for the Poisson equation and the unstable region observed before is only pronounced very slightly. For $m$ between $5$ and $10$ a small instability can be noted, but it does not lead to not achieving the low residual threshold in $100$ iterations. The increase of number of iterations between the various thresholds is larger as before and indicates a slower decrease in residual.
 Comparing the residuals, various QNM can reach with varying degrees of accuracy in the model QLSS, Figure~\ref{fig:Burgers50It} underlines this observation. The residual reached after $50$ iterations is only of order $10^{-7}$ for nearly all $m$ studied. The effect of $m$ for the residual that can be reached in $50$ iterations does not seem to be significant.
 In Figure \ref{fig:BurgersHeatmap} multiple problem sizes were studied for various $m$. One can observe that increasing $m$ does seem to have a very small effect on the number of iterations necessary to reach a residual below $10^{-6}$. The doubling of the problem size seems to increase the iterations necessary by around $10$ to $20$ iterations, while increasing $m$ from $10$ to $20$ does not improve the result. \\
Finally, we have seen that the QNM is capable of solving the highly nonlinear Burgers equation. Unfortunately, compared to classical methods the performance is worse, but a few qubits or in other words relaxed error bounds lead to a stable reduction of the residual. In total, increasing the accuracy of the QLSS does mostly result in guaranteeing convergence rather than improving the speed of it. 

\section{Discussion}
\label{discussion}
The results of this work give new insights into hybrid methods and the application of quantum computers for nonlinear PDEs. In what follows, these findings are critically examined and put into a broader context. Both linear and nonlinear PDEs were solved with the use of the proposed QLSS. The obtained results support the application of the method in the context of CFD. The fact that applying QNM highly accurate approximations were found even with low accuracy QLSS and the increased convergence with increased resources further motivates the application of this hybrid methodology and the potential use on industrial scales. As was observed for the different convergence plots there seems to be the possibility for a trade-off between number of qubits used to encode the eigenvalues and number of iterations needed by Newton's method. Once measurement is also considered this becomes a three way trade-off where the number of qubits, number of shots and the number of Newton iterations can all be interchanged to some degree. This, as has also been shown, has limits and for low accuracy QPE the highly nonlinear problem did not converge at all.\\
To overcome strong computational limitations when it comes to simulating quantum systems, a classical version of the algorithm was deployed. This model QLSS does not represent the full quantum algorithm but gives necessary insights into the scaling behavior of the method and its non-quantum error sources. Nevertheless, these results must be taken with caution since the behavior on quantum hardware could differ due to other sources of error being introduced. Further, since no quantum hardware was accessible there was no possibility to compare the actual runtime of the quantum method. \\
The overall limitations did not allow for any large-scaled problem to be studied, but using current implementations theoretical resource estimations can be made. This estimate is not considering the full stack of quantum hardware, for example error-correction, specific basic gate set decomposition or classical pre-processing. Therefore, this should be considered as a minimal lower bound of necessary resources in regards of depth and width of the circuit. Further, state preparation and measurement, both not considered here, will add another level of complexity to the method.\\
For a large-scale problem $Ax=b$ with $N$ unknowns the algorithm needs $n=\lceil\log_2(N)\rceil$ qubits for the encoding of vector $b$. As mentioned in Section \ref{sec:errorBoundsandAsymScaling} given a desired accuracy of $\varepsilon$ to encode and invert the eigenvalues of the problem matrix we obtain $m = \lceil\log_2(\varepsilon^{-1})\rceil$. The method uses the register $C$, register $M_1$ both of size $m$ and register $M_2$ of size $2m$. Additionally, the current implementation uses two ancilla for the multiplication and comparison operations and one additional ancilla for the $Anc_{flag}$ register. For non-hermitian matrices A, the problem size will be doubled, which results in adding another qubit. Finally, we obtain the following circuit width or total qubit count
\begin{equation}
    \label{ResEstEq}
    Q_{total} = n + 4m + 3 + 1.
\end{equation}
Considering large-scaled problems we assume the asymptotic scaling will dominate over any specific implementations and therefore the circuit depth will be of order $\operatorname{poly}(\kappa, n,\frac{1}{\varepsilon})$.\\
\begin{table}[h]
\centering
\caption{Resource estimations for various number of unknowns $N$ and accuracies $\varepsilon$. The number of qubits $Q_{total}$ is based on Equation \ref{ResEstEq}.}
\label{ResEstTable}
\begin{tabularx}{0.8\textwidth}{X|X|X|X|X}
$N$     & $n$  & $\varepsilon$ & $m$   & $Q_{total}$ \\\hline
$10^{12}$ & 40 & $10^{-12}$      & 80  & 363      \\ \hline
$10^{12}$ & 40 & $10^{-16}$      & 107 & 471      \\ \hline
$10^{16}$ & 54 & $10^{-12}$     & 80  & 377      \\ \hline
$10^{16}$ & 54 & $10^{-16}$      & 107 & 485      \\ \hline
$10^{20}$ & 67 & $10^{-12}$      & 80  & 390      \\ \hline
$10^{20}$ & 67 & $10^{-16}$      & 107 & 498      \\ \hline
$10^{24}$ & 80 & $10^{-12}$      & 80  & 403      \\ \hline
$10^{24}$ & 80 & $10^{-16}$     & 107 & 511     \\\hline
\end{tabularx}
\end{table}
Let us apply the general resource costs to an industrial-scaled use-case in the area of aerodynamics. Considering a Reynolds number of $Re\sim 10^8$, for example to simulate the flow around an airplane in scale-resolving discretization, one would obtain around $Re^{3}=10^{24}$ unknowns in space and time combined \cite{Coleman2010}. Following the estimate by Sanavio and Succi \cite{SanavioSucci.2024} this relates to $\log_2(10^{24})=80$ qubits to encode the full flow field including time on a quantum computer in amplitude encoding. Assuming a desired error of $\varepsilon = 10^{-12}$ and following Table \ref{ResEstTable} we obtain $Q_{total}:=403$ qubits. Given current hardware this seems to be achievable in the near term future. Unfortunately, the total depth does seem to be the main hindrance, QPE being the main contributor with a circuit depth for this use-case of order $10^{12}$. To be able to compute circuits of such depth one needs fault-tolerant hardware not existing at the moment.\\
The full procedure would then again embed a QLSS of this size into the QNM and the question arises how many times the QLSS would be called. It is not an easy task to estimate the necessary iterations needed by Newton's method for problems of this magnitude. Looking at successful implementations of Newton's methods for smaller test cases of the order of $Re\sim 10^6$ using simplified equations one can obtain convergence after around $5000$ to $10000$ iterations \cite{La:18}. In general, one can expect these methods to converge in less iterations than time-marching schemes, iterating over all the time steps and solve the PDE for each such time step. Therefore this method is an improvement over such hybrid time-stepping methods, which communicate between classical and quantum systems after every time step or only obtain the final state, as were proposed for example by \cite{Over.10.10.2024Vortex, WAWRZYNIAK2025_QLBM_timemarching}.\\
Accessing the full exponential advantage in practice poses a difficult challenge. For this the full end-to-end application of the method must be considered. This mainly encompasses an efficient way to communicate between classical and quantum systems, i.e. a state preparation and measurement preserving the computational speed up of the quantum system. Due to the physical structure of the solutions considered here, it seems reasonable that we can improve on brute force state preparation and measurement, which would make the proposed method lose any advantage. For the state preparation some work has been done keeping a sublinear complexity, but not an exponential speed up \cite{pagni2025sublinearclassicaltoquantumdataencoding}. Additional work into various state preparations has been done \cite{Araujo.2021, Zhang.2022StatePrep}. To obtain the information from the quantum system one could propose to run the algorithm exponentially often in parallel, which quickly would become infeasible especially for the industrial use-cases. Some measurement strategies exist, which take advantage of underlying structure through neural networks \cite{Torlai.2018} or matrix product state representations \cite{Cramer.2010} and need to be studied further for the proposed method. A cut-off measurement, i.e. only measuring the biggest values in $\Delta u$ has been studied for this special application, but the obtained runtime has not been satisfactory \cite{Patterson.2025}. Further, it would be necessary to enhance these strategies with amplitude amplification to improve the success probability \cite{Brassard_2002} and special measurement strategies to obtain the relative signs have to be deployed \cite{Manzano.2023RealAmpEst}. Apart from the challenge to provide efficiently designed state preparation and measurement procedures, in practice $A$ is not always well-conditioned, increasing the computational complexity of the algorithm. In current CFD methodology this also poses a problem and therefore so called preconditioners get applied to improve the condition number of the matrices, see for example \cite{La:18}. Additionally, preparation of the circuit, which depends on the derived linear system to be solved, needs to be possible without eliminating the computational benefits. Therefore an efficient scheme to obtain quantum circuits for the Hamiltonian simulation must be implemented.\\

\section{Conclusion}
\label{conclusion}
In this paper we first presented a new variant of the HHL algorithm, which reduces the prior knowledge of the eigenvalues of the matrix one wants to invert. The method is showcased on a linear PDE problem and on randomly generated matrices. Afterwards, the QLSS is applied in a hybrid classical-quantum method to solve nonlinear PDEs using Newton's method. The QNM is applied in order to approximately solve a nonlinear Poisson equation and the Burgers equation. The performance was studied using Qiskit simulations as well as classical versions of the algorithm. Even though the hybrid method has been previously proposed, relatively little work was done to showcase its practical applicability. Using current implementations, a resource estimation for industry-scale use-cases is provided.\\
The novel HHL variant performs well on various random matrices, as well as on linear systems arising in solving linear PDEs. By increasing the number of qubits and the circuit depth, the algorithm improves and can solve linear systems with any desired accuracy. This paper provides insights how Newton's method coupled with quantum methods is capable of solving nonlinear PDEs. The results showcased here are limited in size due to computational limitations but hint at an applicability for potential CFD problems. It still requires multiple state preparations and read out steps, both of which are not straight forward tasks under computational speed-up considerations.\\
This paper is supposed to serve as a proof of concept with future work focusing on a full procedure considering state preparation and measurement strategies. It would be worthwhile to design more advanced circuit implementations to reduce depth. The width of the current circuit has been shown to be less prohibitive than the depth. This opens the possibility to use ancilla qubits to shorten the circuit. It further demonstrates, that the demand on the quantum hardware is not solely focused on the number of qubits, but also on the accuracy of the operations, longer decoherence times and an efficient communication between systems. Additionally, the algorithm needs to be generalized to handle negative eigenvalues. Combining this method further with standard CFD practices such as preconditioning, multi-grid or variants of Newton's methods might also benefit the applicability of this method.\\ 

\section*{Acknowledgments}
This project was made possible by the DLR Quantum Computing Initiative and the Federal Ministry for Economic Affairs and Climate Action;
\url{qci.dlr.de/projects/toquaflics}

\bibliographystyle{siamplain}
\bibliography{literature}
\end{document}